\documentclass[]{aastex631}

\usepackage{graphicx}
\usepackage{subfigure}
\usepackage{color}
\usepackage{gensymb}
\usepackage{ftnxtra}
\usepackage{fnpos}

\shorttitle{Re-analysis on WASP-35 and HAT-P-30 systems}
\shortauthors{Lu Bai et al.}
\graphicspath{{./}{figures/}}

\begin{document}

\title{WASP-35 and HAT-P-30/WASP-51: re-analysis using TESS and ground-based transit photometry}

\author{Lu Bai}\email{Corresponding author: wangxb[at]ynao[dot]ac[dot]cn (XBW), sunleile[at]ynao[dot]ac[dot]cn (LLS)}\thanks{Lu Bai and Leilei Sun contributed equally to this work.}
\affiliation{Yunnan Observatories, Chinese Academy of Sciences, Kunming 650216, China}
\affiliation{Key Laboratory for the Structure and Evolution of Celestial Objects, Chinese Academy of Sciences, Kunming 650216, China}
\affiliation{University of Chinese Academy of Sciences, Beijing 100049, China}

\author{Shenghong Gu}
\affiliation{Yunnan Observatories, Chinese Academy of Sciences, Kunming 650216, China}
\affiliation{Key Laboratory for the Structure and Evolution of Celestial Objects, Chinese Academy of Sciences, Kunming 650216, China}
\affiliation{School of Astronomy and Space Science, University of Chinese Academy of Sciences, Beijing 101408, China}

\author{Xiaobin Wang}
\affiliation{Yunnan Observatories, Chinese Academy of Sciences, Kunming 650216, China}
\affiliation{Key Laboratory for the Structure and Evolution of Celestial Objects, Chinese Academy of Sciences, Kunming 650216, China}
\affiliation{School of Astronomy and Space Science, University of Chinese Academy of Sciences, Beijing 101408, China}

\author{Leilei Sun}\thanks{Lu Bai and Leilei Sun contributed equally to this work.}
\affiliation{Yunnan Observatories, Chinese Academy of Sciences, Kunming 650216, China}
\affiliation{Key Laboratory for the Structure and Evolution of Celestial Objects, Chinese Academy of Sciences, Kunming 650216, China}

\author{Chi-Tai Kwok}
\affiliation{Ho Koon Nature Education cum Astronomical Centre, Sik Sik Yuen, Hong Kong, China}

\author{Ho-Keung Hui}
\affiliation{Ho Koon Nature Education cum Astronomical Centre, Sik Sik Yuen, Hong Kong, China}



\begin{abstract}

High-precision transit observations provide excellent opportunities for characterizing the physical properties of exoplanetary systems. These physical properties supply many pieces of information for unvealing the internal structure, external atmosphere, and dynamical history of the planets. We present revised properties of transiting systems WASP-35 and HAT-P-30/WASP-51 through analyzing newly available TESS photometry and ground-based observations obtained at 1m telescope of Yunnan Observatories as well as from the literature. The improved system parameters are consistent with the previous results. Furthermore, we find that HAT-P-30b/WASP-51b's transits show significant timing variation which cannot be explained by decaying orbit due to tidal dissipation and the R\o mer effect, while both apsidal precession and an additional perturbing body could reproduce this signal through our comprehensive dynamical simulations. Because both of them are valuable targets which are suitable for transmission spectroscopy, we make some predictions for atmospheric properties of WASP-35b and HAT-P-30b/WASP-51b based on newly derived system parameters.

\end{abstract}



\section{Introduction} \label{sec:intro}

Transiting exoplanets have been playing a fundamental role in revolutionizing the planetary science for more than two decades. High precision photometry during the transits and spectroscopic measurements provide perfect opportunities to characterize the physical properties of transiting exoplanetary systems. These information give the first hints about the internal structure, external atmosphere, and dynamical history of the planetary systems. Therefore, it is important to enlarge the number of the exoplanetary systems with accurately measured physical properties aiming to the statistically analyze as well as establish the formation and evolution models of planetary systems.

Long-term transit monitoring observations and thus physical properties updating for known transiting exoplanetary systems are essential for discovering additional bodies and scheduling further observations. A single transiting planet generally orbits around its host star on a Keplerian orbit with a constant orbital period in the timescale of years, in which the influence of tidal effect and general relativity effect can be negligible due to no effective accumulation. But the transit timing variations \citep[TTV;][]{RN323,RN684}, which means transits no longer appear at a fixed interval, would exist in principle if there are additional bodies in the planetary systems. In some cases the gravitational interaction between planets will trigger relatively short-term TTVs \citep{RN323,RN684}. The patterns of these TTVs (e.g., amplitude, frequency, and overall shape) strongly rely on the orbital parameters and masses of the planets involved \citep[see e.g.][]{RN323,RN684,RN719,RN313,RN720}. As the gravitational interactions among planets that induce the TTVs generally act on timescales much longer than the orbital periods, space-based transit survey missions, such as {\it Kepler} and upcoming PLATO, and ground-based follow-up transit monitoring with longer baselines are more likely to capture such phenomena \citep{RN721,RN722,RN723}. Furthermore, TTV technique is a powerful tool for understanding planetary systems: it can place a constraint on the existence of non-transiting exoplanets, thereby remedying missing pieces to the architecture of the systems due to the geometry bias which is inherent to the transit method \citep{RN720,RN724,RN725}, and allowing for a better comparison with synthetic planetary system population models \citep[see e.g.][]{RN726,RN727,RN789}. On the other hand, TTVs can also be used to measure the masses of the planets \citep[see e.g.][]{RN728,RN313}, and therefore their density, which hence place strong constraints on their internal structures, such as the cases for Kepler-411 system \citep{RN685}, Trappist-1 system \citep{RN729,RN730} and so on. Detection of individual dynamically hot systems also provides valuable constraints on planetary system formation theory, as the current orbit of a system can remain the information of its past migration history \citep[see e.g.][]{RN731,RN732}. 

In addition, there are many exoplanetary systems show temporal variations in their orbital properties, several reasons may cause the change in orbital properties, such as the long-term effects of tidal forces, the elliptical orbit precession, mass loss , and so on; these mechanisms may mimic above mentioned TTVs generated by the gravitational interactions with other bodies. Among these cases, orbital decay induced by tidal dissipation is another most interesting one besides the gravitational interactions with other bodies, relative theory suggests it tends to occur in planets with shorter orbital periods and larger masses; such kind of TTV signals of hot giant planets could also be used to constrain the planetary tidal quality factor and study the dynamical history of planetary systems \citep{RN269}. While, direct observational evidences still remain sparse, further investigations will improve our understanding on the dynamical history of planetary systems. Fortunately, both space- and ground-based transit photometry can be used to search for TTVs, this will accumulate a lot of data and valuable targets.

To refine physical properties and investigate the TTV behivour of known transiting exoplanetary systems, both ground- and space-based transiting light curves are necessary. Since 2009, we have monitored the transit events of some known transiting exoplanetary systems by employing the 1m telescope of Yunnan Observatories (hereafter, YO-1m) in China, and obtained a series of high-precision photometric data \citep[see e.g.][]{RN747,RN338,RN748,RN383,RN319}. For the ground-based transit observations, the most challenging thing is the systematic noise due to the variable atmosphere of the Earth, so some noise reduction techniques \citep{RN377,RN378,RN382,RN524,RN525,RN783} have been developed and widely used to handle this issue. Moreover, the Transiting Exoplanet Survey Satellite \citep[TESS;][]{RN697,RN514} has been producing high quality transiting light curves which are well suited for refining the physical properties of known planetary systems. 

Here we focus on the planetary systems WASP-35 and HAT-P-30. Both of them had been observed by TESS in two sectors, and we also observed one transiting event for each system using YO-1m. Based on these  light curves and radial velocity (RV) data from the literature, we refined their system parameters, analyzed the TTVs of these planetary systems, and made predictions about the atmospheric properties of them based on our refined physical parameters. We give a short introduction of the targets in Section \ref{sec:tar}. Then we describe the observations and data reduction in Section \ref{sec:obs}, and present the modeling in Section \ref{sec:tra}. The further analysis and discussion is given in Section \ref{sec:dis}. At last, we summarize the study in Section \ref{sec:con}. 

\section{Targets}\label{sec:tar}

\subsection{WASP-35}

WASP-35b was discovered by \citet{RN516} in the Wide Angle Search for Planet (WASP) project. This transiting exoplanet system has a inflated ($M_{p}=0.72\pm0.06 M_{Jup}$ and $R_{p}=1.32\pm0.05 R_{Jup}$) hot Jupiter which orbits a metal-poor ($[Fe/H]=-0.15\pm0.09$) host star with a period of 3.16 days. Through analyzing their photometric and spectroscopic observations at a series of ground-based telescopes, \citet{RN516} confirmed the planetary nature of WASP-35b and suggested that the host star of WASP-35 was lack of stellar activity. Later, \citet{RN519,RN520} updated the physical parameters of the host star using a spectroscopic analysis, \citet{RN784} re-analyzed and updated the ephemeris based on the light curves of \citet{RN516}, and \citet{RN785} refined and updated the ephemeris using the data of TESS photometry.

\subsection{HAT-P-30}

HAT-P-30b (also known as WASP-51b) was discovered independently by \citet{RN518} in the Hungarian-made Automatic Telescope Network (HATNet) and \citet{RN516} in the WASP project, respectively. HAT-P-30b is a transiting hot Jupiter ($M_{p}=0.711\pm0.028 M_{Jup}$ and $R_{p}=1.340\pm0.065 R_{Jup}$) orbiting a late F-type host star with a period of 2.81 days and the orbit is circular. HAT-P-30 has a relatively bright (V=10.4) host star, therefore it is very suitable for using transmission spectra to study the planetary atmosphere. \citet{RN518} studied the Rossiter-McLaughlin effect of HAT-P-30 system, and found the orbit of HAT-P-30b was highly tilted with a sky-projected angle between the star's spin axis and the planet's orbit normal, $\lambda=73.5\pm9.0\degr$. \citet{RN517}, \citet{RN699}, \citet{RN700} and \citet{RN786} acquired new transit light curves of HAT-P-30b based on several ground-based telescopes and measured the physical parameters. \citet{RN698} analyzed several relative high quality transit light curves of HAT-P-30b collected from the Exoplanet Transit Database (ETD) \footnote{http://var2.astro.cz/ETD/}. \citet{RN784} re-analyzed and updated the ephemeris based on the light curves of \citet{RN517}, \citet{RN518} and \citet{RN516}, and found no TTVs for HAT-P-30b.

\section{Observations and data reduction} \label{sec:obs}

\subsection{TESS photometry}

TESS is an all-sky space survey, which is designed to search for transiting exoplanets orbiting the bright and nearby stars, these planet systems are enable us to carry out the follow-up observations and study their physical properties. TESS performs time-series photometry to monitor at least 200,000 main-sequence stars using four 100 mm telescopes with wide-field optical CCD cameras. The broad passband filter of TESS covers from 600 to 1000 nm and the combined field of view of each sector is 24 $\times$ 96 square degrees. Since its launch in 2018, TESS has completed 44 sectors of observations and collected a lot of high-quality transiting light curves. 

\begin{figure}
   \centering
   \includegraphics[width=\hsize]{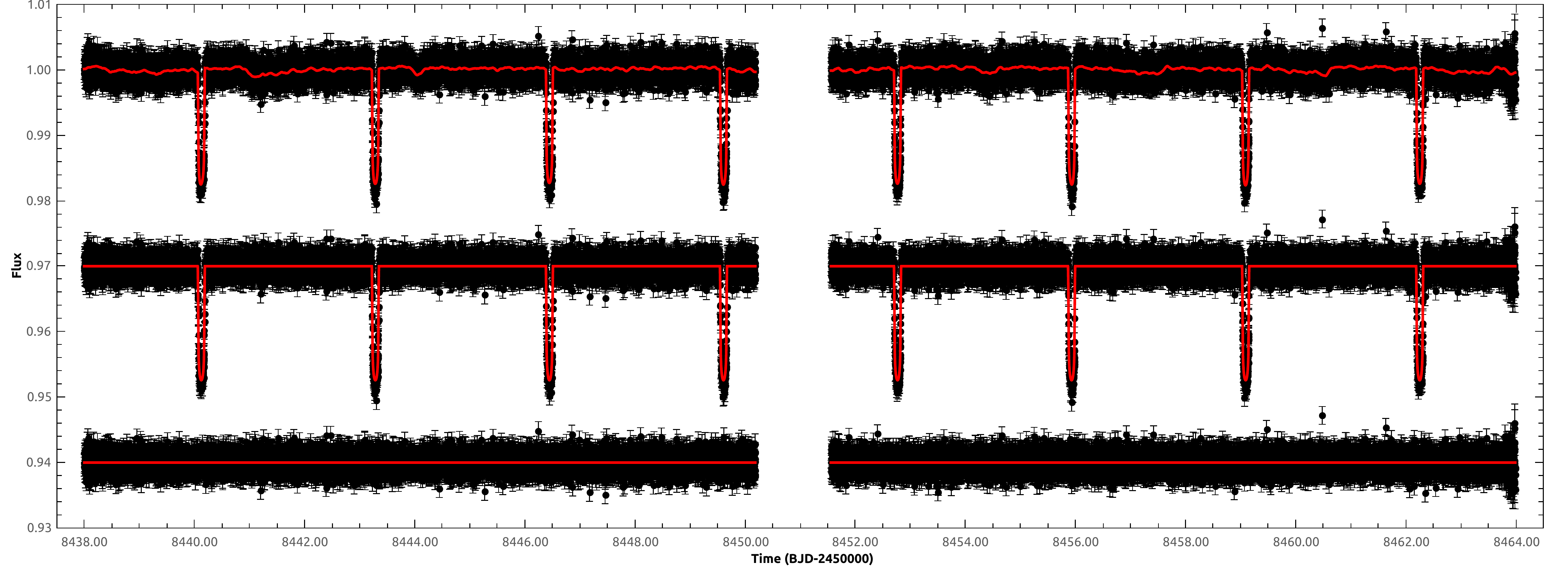}
      \caption{Transit light curves of WASP-35 observed by TESS in Sectors 5. The top is raw light curves with the best-fitting transit + noise model, the middle is the detrended light curves with the best-fitting transit model, and the bottom is the corresponding residuals. Vertical shifts are added for visualization.}
         \label{fig1}
   \end{figure}

\begin{figure}
   \centering
   \includegraphics[width=\hsize]{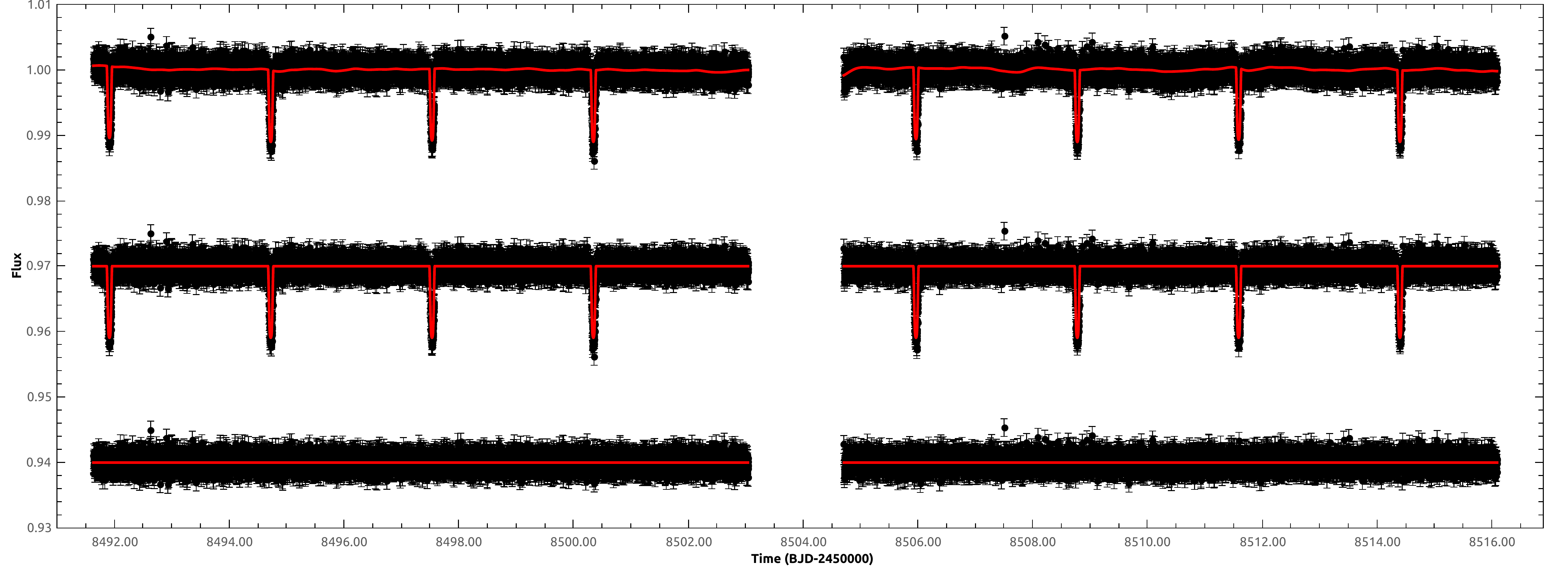}
      \caption{Transit light curves of HAT-P-30 observed by TESS in Sectors 7. The top is raw light curves with the best-fitting transit + noise model, the middle is the detrended light curves with the best-fitting transit model, and the bottom is the corresponding residuals. Vertical shifts are added for visualization.}
         \label{fig2}
   \end{figure}

WASP-35 was observed by TESS in Sectors 5 (2018 November 15 - 2018 December 11) and Sectors 32 (2020 November 19 - 2020 December 17), and HAT-P-30 was observed by TESS in Sectors 7 (2019 January 7 - 2019 February 2) and Sectors 34 (2021 January 13 - 2021 February 9). The data were collected in a 2 minute cadence and reduced by the pipeline developed by the Science Processing Operations Center \citep[SPOC;][]{RN702}. We downloaded the light curve files from the archives at Mikulski Archive for Space Telescopes (MAST)$\footnote{https://archive.stsci.edu/mission/tess/}$ and accessed the light curves using the \textbf{Lightkurve} python package \citep{RN521}. These data had been corrected for the instrumental systematic variations by using the SPOC pipeline, but some long-term trends still remained in the data. In order to alleviate the influences from the remaining trends, we employed Gaussian Process \citep[GP;][]{RN524} to model these data (see below for further details). At last, we obtained 15 complete transit light curves of WASP-35 and 16 complete transit light curves of HAT-P-30. The time system of these light curves is barycentric Julian date (BJD) which can be used directly to determine the system parameters and ephemerides. The examples of raw TESS light curves are displayed in the top panels of Figure \ref{fig1} and Figure \ref{fig2}, respectively.

\subsection{Ground-based photometry}

\begin{figure}
   \centering
   \includegraphics[width=\hsize]{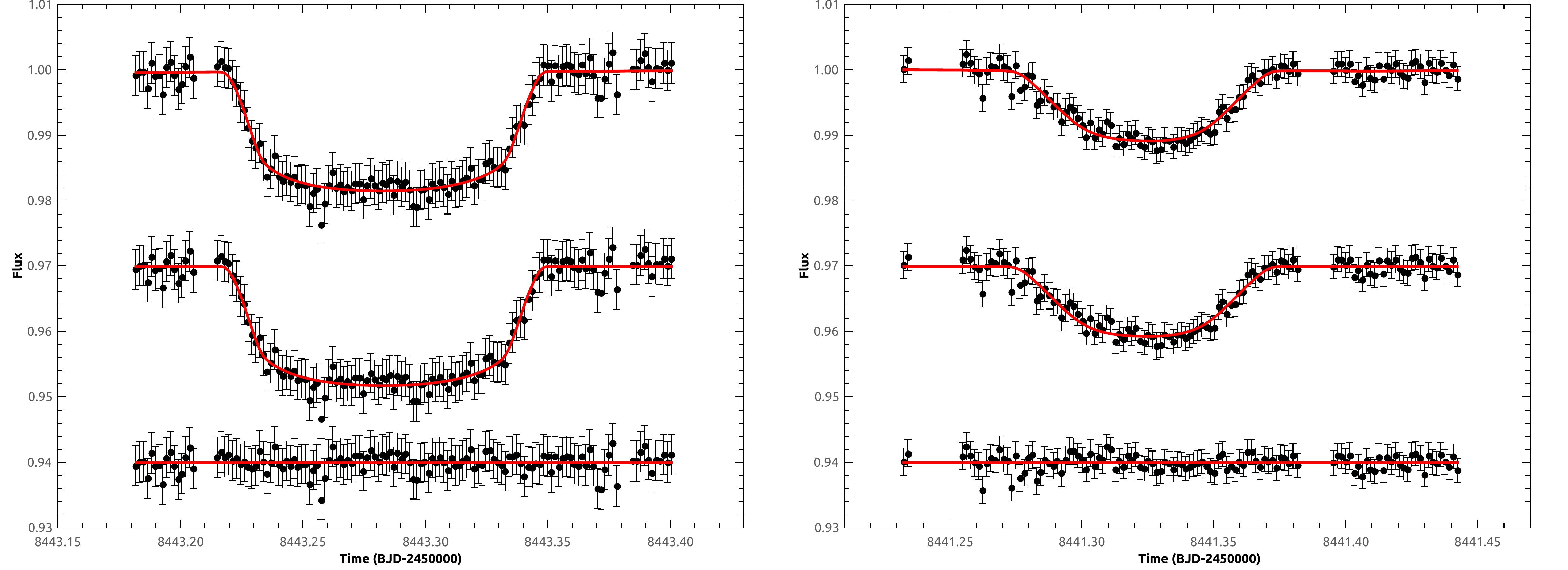}
      \caption{The transit light curves of WASP-35 (left panel) and HAT-P-30 (right panel) observed using YO-1m. The top row is raw light curves with the best-fitting transit + noise model, the middle row is light curves removed the noise with the best-fitting transit model, and the bottom row is the corresponding residuals.}
         \label{fig3}
   \end{figure}

We observed the transit events of WASP-35 and HAT-P-30 using the Andor 4K$\times$4K CCD camera attached to the YO-1m on 2018 November 20 and 18, respectively.  The Johnson-Cousins R filter was utilized in both observations and the field of view is 15$\times$15 arcmin$^{2}$. During two observations, both weather conditions and instrument statuses were good, and the exposure times were 120s. The transit event of HAT-P-30b occurred at sunset, which resulted in a relatively larger dispersion at the beginning of the observation (see Figure \ref{fig3}). In addition, we collected two sets of high quality light curves of HAT-P-30 from the CDS database, these light curves were observed by \citet{RN518} using the KeplerCam imager of the 1.2 m telescope at the Fred Lawrence Whipple Observatory (FLWO-1.2m).  

The data reduction follows the standard procedures which were described in \citet{RN338}. Our data reduction pipeline is based on the IRAF package, written in the python language by means of the pyRAF interface, including image checking, image trimming, bias subtraction, flat-field correction, cosmic ray removal, establish astrometric solution, aperture photometry  and systematic error correction. To reach a higher precision, the pipeline attempts a series of photometric apertures for the target and reference stars and we pick out an optimal one to minimize the dispersion of each light curve. The transit light curve is obtained by using the optimal aperture.

For the ground-based observations, the shallow transit signals are easy to be drowned out by the systematic errors due to the effect of the telluric atmosphere and imperfect observation conditions, so it is necessary to remove the systematic errors hiding in above transit light curves. First, we employed the coarse de-correlation \citep{RN377} and SYSREM algorithms \citep{RN378} to diagnose and correct them by using the pipeline mentioned above. At this point, there are still some long-term systematic trends in the transit light curves, we employed GP to diagnose and correct them. Furthermore, in order to obtain accurate ephemerides of targets, we converted the observing time into barycentric Julian date by using the method proposed by \citet{RN379}. 

\begin{figure}
   \centering
   \includegraphics[width=\hsize]{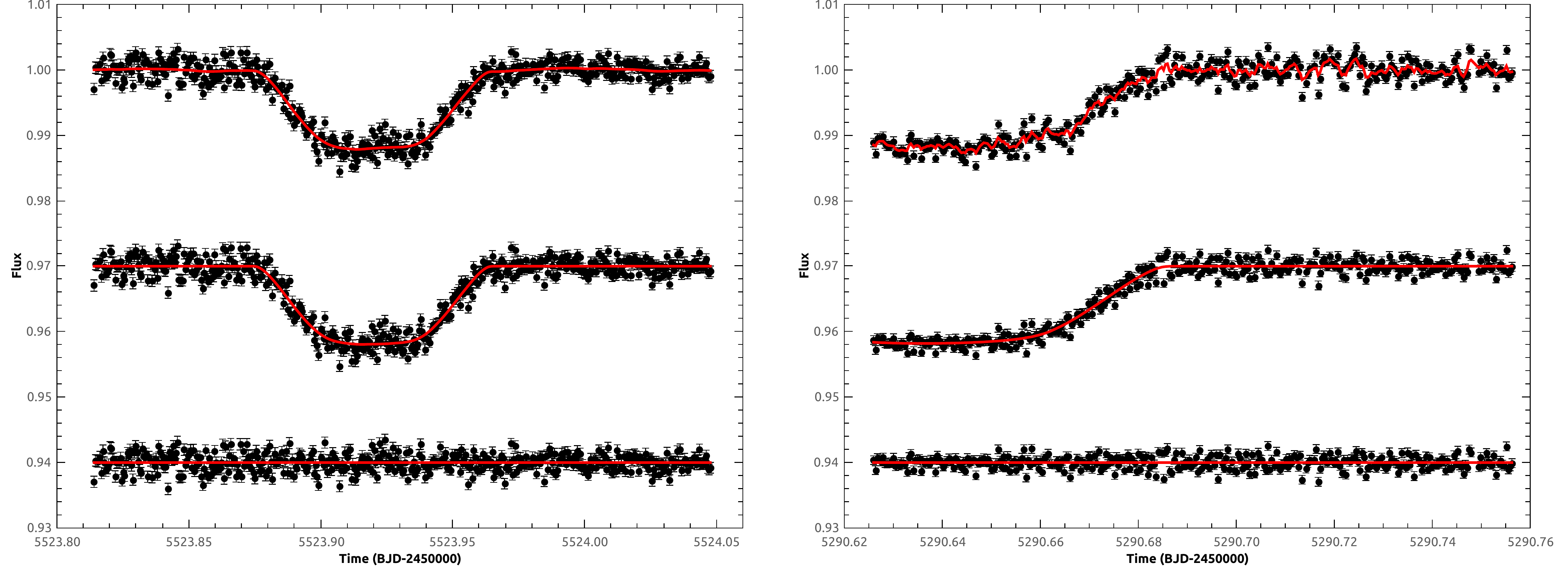}
      \caption{The transit light curves of HAT-P-30 observed using FLWO-1.2m. The top row is raw light curves with the best-fitting transit + noise model, the middle row is light curves removed the noise with the best-fitting transit model, and the bottom row is the corresponding residuals.}
         \label{fig4}
   \end{figure}

\subsection{Gaussian Processes}\label{subsec:gp}

Gaussian Process \citep{RN524} is a popular method which has been widely used to model time series data in the exoplanet community for about a decade, such as modeling stellar activity signals in RV data and correcting instrumentally induced systematic errors in transit light curves \citep{RN525,RN526,RN703,RN704,RN705}. Instead of modeling instrumental systematic errors as a deterministic function with auxiliary measurement parameters, GP provides a non-parametric method to model systematic errors from the observed data set. Here, we used the code of \textbf{juliet} \citep{RN527} to corrected long-term trends and instrumental systematic errors in TESS photometry data and the residual systematic errors in the ground-based photometry data.

We modeled the photometric light curves in flux space separately depending on the instrument and the time of the observation by using \textbf{juliet} code. For the photometry data sets, \textbf{juliet} establishes a common model in which light curve is modeled as a linear combination of the transit model and the noise model:
\begin{displaymath}
		\mathcal{M}(t)+\epsilon(t)
\end{displaymath}
where $\mathcal{M}(t)$ is the transit model with the dilution factor for the given instrument and the mean offset of out-of-transit flux, $\epsilon(t)$ is the noise model which is being modeled by GP in the photometric light curve. \textbf{Juliet} uses the analytic transit light curve model of \citet{RN381} with a quadratic limb-darkening law and the limb-darkening coefficients ($q_{1}$ and $q_{2}$) proposed by \citet{RN529} to model the transit signals and employs \textbf{batman} package \citep{RN528} to implement the transit light curve model. 

We checked the nearby stars around WASP-35 and HAT-P-30 using Gaia DR3 database \citep{RN515}, both of the fields do not seem to be especially crowded, and there was no other source bright enough to make a perceptible impact, so we fixed the dilution factors of both of the targets to 1. For TESS photometry data, the state of the instrument may be different in different observation sectors, so we modeled the TESS data sector by sector, respectively. For the ground-based observations, the state of ground-based instrument and the weather is changing from day to day, so we modeled the ground-based data day by day, respectively. In all of the model processes, we chose a celerite (approximate) Matern multiplied exponential kernel within \textbf{Juliet} implemented by \textbf{celerite} package \citep{RN530}. For the analytic transit model part,  the parameters are the orbital period $P$, the mid-transit time $T_{0}$, scaled semi-major axis $a/R_{A}$, the eccentricity $e$ (fixed to 0), the argument of periastron $\omega$ (fixed to 90 degrees), the impact parameter $b$, the planet-to-star radius ratio $R_{b}/R_{A}$, the limb-darkening coefficients $q_{1}$ and $q_{2}$, the mean out-of-transit flux, the dilution factors (fixed to 1), and the jitter. The input values we used were obtained by \citet{RN516} for WASP-35 and \citet{RN518} for HAT-P-30 respectively, and a wide normal prior was used in the GP. For the GP component, the hyperparameters were the amplitude of the GP, two length scales corresponded to the Matern and the exponential part. We used a log-uniform prior for these hyperparameters, the amplitude of the GP varied from $10^{-6}$ to $10^{6}$, and both of the two length scales varied from $10^{-3}$ to $10^{3}$. \textbf{Juliet} uses \textbf{PyMultiNest} package \citep{RN532,RN531} with 500 live points to explore the parameter space. Finally, we obtained the raw light curves and the full median posterior model, namely the transit model plus the median GP process. So we could subtract the GP part of the model from the raw light curve to correct the residual systematic errors in photometry data and obtained the final light curves.  The transit light curves before and after the GP correction with the best-fitting models and the residuals are shown in Figure \ref{fig1}, Figure \ref{fig2}, Figure \ref{fig3} and Figure \ref{fig4}. 

\section{Transit and RV Modeling}\label{sec:tra}

We used the Markov Chain Monte Carlo (MCMC) technique to fit the final light curves and the RV curves of \citet{RN516} and \citet{RN518} simultaneously, and derived the system parameters as well as the ephemerides for further analysis \citep{RN533,RN534}. The code which we employed to model the transit light curves and RV curves was developed by \citet{RN533}, which is based on the analytic transit light curve model proposed by \citet{RN381} associating with the four-coefficient limb-darkening law. The input parameters include: the orbital period $P$, the mid-transit time $T_{0}$, the transit duration $t_{T}$, the planet to star area ratio $\Delta F$, the impact parameter $b$, the semi-amplitude of the RV curve $K_{1}$, the orbital eccentricity $e$ and the argument of periastron $\omega$. In order to accelerate the convergence of the system parameters, $e\cos\omega$ and $e\sin\omega$ are used to replace $e$ and $\omega$ in the MCMC processes. The mass and radius of host stars were derived from the calibration for stellar masses and radii proposed by \citet{RN537}, which is based on the stellar effective temperature, metallicity and density. The limb-darkening coefficients were obtained by interpolating the coefficient tables of \citet{RN535,RN536} according to the stellar effective temperature, metallicity, surface gravity and microturbulent velocity of host stars. The stellar parameters mentioned above were based on the results of  \citet{RN516} and \citet{RN518}, and we adopted a circular orbit ($e=0$) in following analysis of both of targets, as both \citet{RN516} and \citet{RN518} did not find a significant non-zero eccentricity through fitting their RV curves. The Metropolis-Hastings algorithm was used to obtain the posterior probability distribution of the system parameters and the best-fit system parameters with uncertainties.

At the beginning of the analysis, we modeled all of the light curves and the RV curves simultaneously to derive the initial global system parameters of the planetary systems. The 12 RV measurements of WASP-35 were made by \citet{RN516} between 2010 January 5 and 2010 February 14, and the RV curves of HAT-P-30 were observed by \citet{RN516} and \citet{RN518} between 2010 April 27 and 2011 January 4, including 39 data points. The inputs parameters were described above, we adopted a circular orbit ($e=0$) and other system parameters as free ones varied with the MCMC calculations. We employed 5 chains of 17,000 MCMC steps, in which 2,000 steps were burn-in samples and had been eliminated in the statistics of posterior probability distribution. This chain length was carefully considered, which had balanced convergence of the solution and computation time \citep{RN338}. 

Then we fitted the separately to derive the mid-transit times for the TTV analysis and refining the ephemerides, using the above results as the input parameters. Other configurations followed the same procedure as above, except that the orbital periods were fixed as the input ones. The results of the mid-transit times of the targets are listed in Table \ref{tab1} and Table \ref{tab2}, combined with several mid-transit times from previous works and the ETD website. we used a linear ephemeris formula $T = T_{0} \times E$ to fit the new mid-transit times and derived the orbital period values respectively, where $T_{0}$ is the zero point of the epoch of mid-transit, $P$ is the orbital period and $E$ is the the cycle number. we preformed the MCMC calculations to find the best-fitting linear ephemeris formula by using the \textbf{emcee} package \citep{RN263}. We ran 50,000 MCMC steps with 1,000 burn-in steps to ensure convergence and the final results are $T(\mathrm{BJD_{TDB}}-2450000) = 5531.47909(15) + 3.1615692(2) \times E$ for the WASP-35 system and $T(\mathrm{BJD_{TDB}}-2450000) = 5523.92157(48) + 2.8106006(4) \times E$ for the HAT-P-30 system.

\begin{figure}
   \centering
   \includegraphics[width=\hsize]{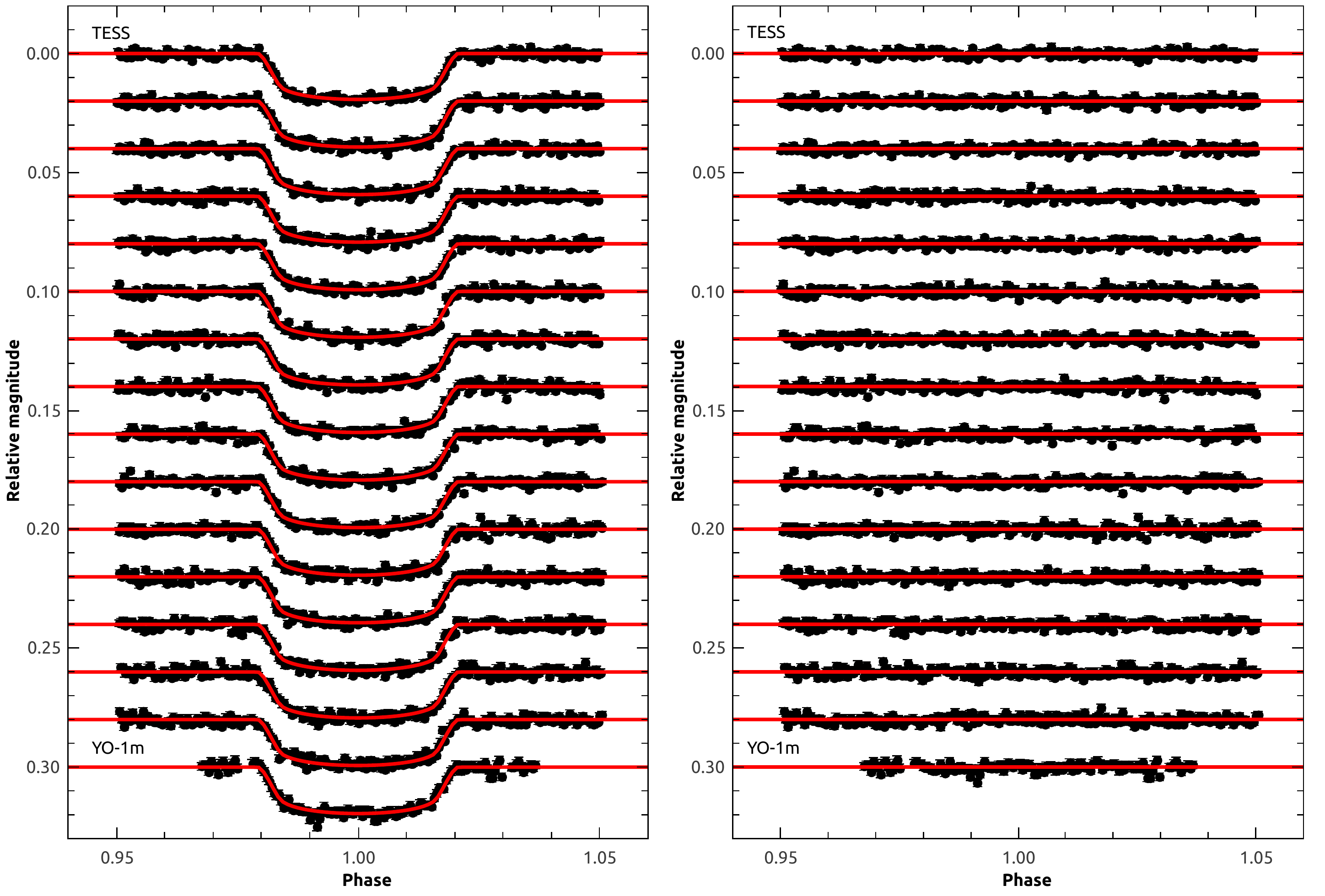}
      \caption{The final transit light curves of WASP-35 with the best-fitting transit models and the residuals.}
         \label{fig5}
   \end{figure}

\begin{figure}
   \centering
   \includegraphics[width=\hsize]{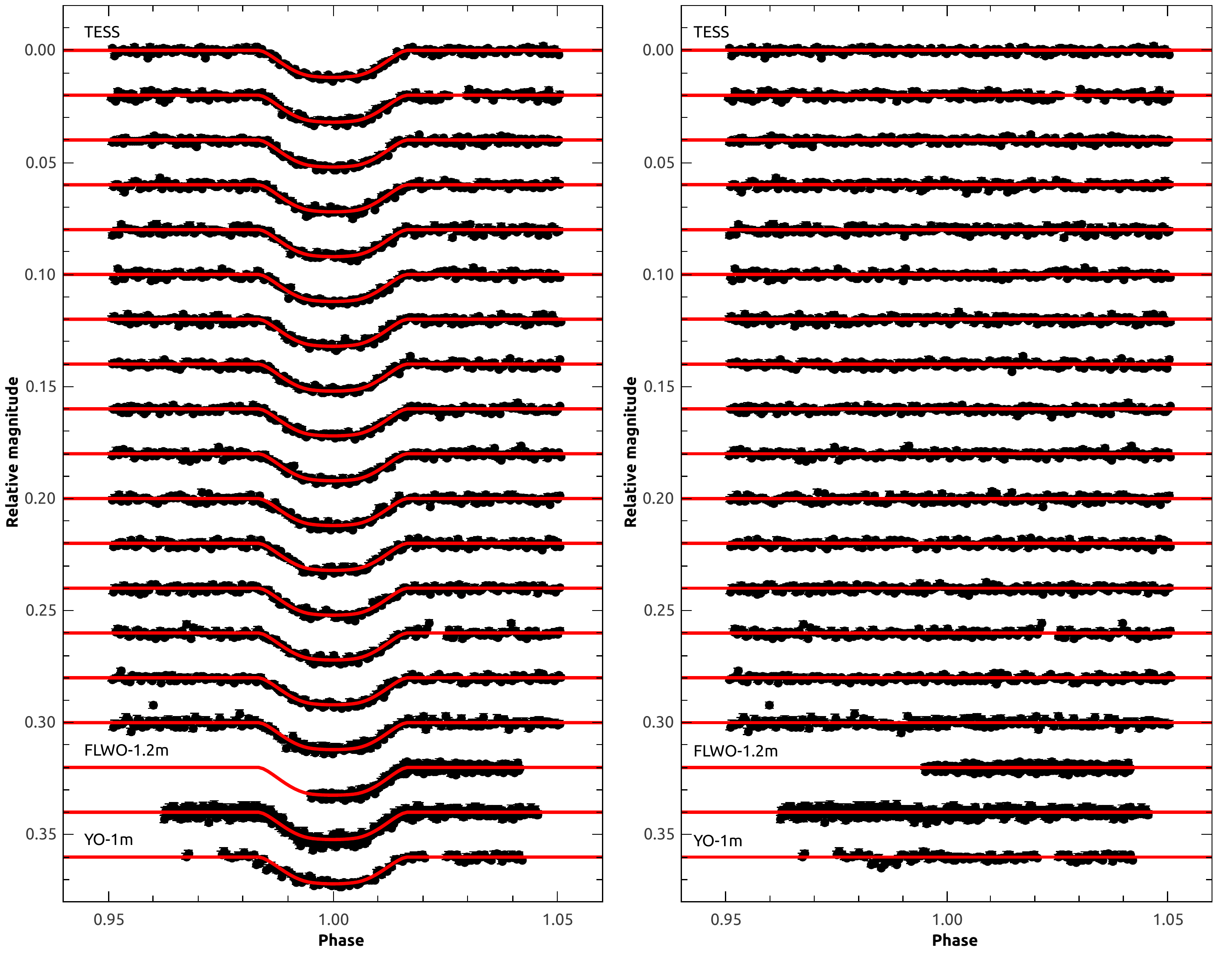}
      \caption{The final transit light curves of HAT-P-30 with the best-fitting transit models and the residuals.}
         \label{fig6}
   \end{figure}

Using the new orbital periods derived above as the input parameters and keeping fixed, the final system parameters of the targets were calculated based on all the light curves and the RV curves following the same strategy as above. The final solutions of the system parameters for WASP-35 and HAT-P-30 systems are listed in Table \ref{tab3} and Table \ref{tab4} together with the results of \citet{RN516} and \citet{RN518} for comparison. The final transit modeling results are shown in Figure \ref{fig5} and Figure \ref{fig6}.

\begin{table}
\caption{The mid-transit times of WASP-35.}             
\label{tab1}      
\centering          
\begin{tabular}{c c c c}  
\hline\hline       
Mid-time & Error & Cycle number & Source \\
($\mathrm{BJD_{TDB}-2450000}$) & (days) & &  \\
\hline
5531.4790700&0.0001500&0&\citet{RN516}\\
8440.1231974&0.0002673&920&TESS\\
8443.2834342&0.0003904&921&YO-1m\\
8443.2845042&0.0002821&921&TESS\\
8446.4460079&0.0002735&922&TESS\\
8449.6077501&0.0003029&923&TESS\\
8452.7689812&0.0002756&924&TESS\\
8455.9305639&0.0002782&925&TESS\\
8459.0921595&0.0002948&926&TESS\\
8462.2537745&0.0003173&927&TESS\\
9157.8000120&0.0008600&1147&ETD\\
9176.7683243&0.0003100&1153&TESS\\
9176.7697320&0.0007200&1153&ETD\\
9179.9301275&0.0002882&1154&TESS\\
9183.0917244&0.0003045&1155&TESS\\
9189.4143000&0.0003028&1157&TESS\\
9192.5756463&0.0003310&1158&TESS\\
9195.7378575&0.0003504&1159&TESS\\
9195.7389020&0.0006300&1159&ETD\\
9198.8988431&0.0003170&1160&TESS\\
9252.6467690&0.0009600&1177&ETD\\
9524.5411270&0.0006800&1263&ETD\\
\hline  
\hline 
\end{tabular}
\end{table}

\begin{table}
\caption{The mid-transit times of HAT-P-30.}             
\label{tab2}      
\centering  
\begin{tabular}{c c c c}  
\hline\hline       
Mid-time & Error & Cycle number & Source \\
($\mathrm{BJD_{TDB}-2450000}$) & (days) & &  \\
\hline
5290.6368010&0.0043224&-83&\citet{RN518}\\
5523.9198071&0.0003303&0&\citet{RN518}\\
5650.3954100&0.0030500&45&ETD\\
5928.6491100&0.0009000&144&ETD\\
5945.5120500&$^{+0.0005400}_{-0.0004900}$&150&\citet{RN517}\\
5945.5136000&0.0019300&150&ETD\\
5970.8068390&0.0016200&159&ETD\\
5976.4282790&0.0012500&161&ETD\\
5976.4288790&0.0011800&161&ETD\\
6240.6257120&0.0012300&255&ETD\\
6296.8385100&0.0011600&275&ETD\\
6679.0801200&0.0013400&411&\citet{RN700}\\
6704.3729040&0.0011600&420&ETD\\
7449.1875100&0.0018300&685&\citet{RN700}\\
7460.4253570&0.0011000&689&ETD\\
7775.2132600&0.0008300&801&\citet{RN700}\\
8140.5896250&0.0013300&931&ETD\\
8143.4050250&0.0012000&932&ETD\\
8157.4556750&0.0013400&937&ETD\\
8171.5092660&0.0015600&942&ETD\\
8441.3230170&0.0007651&1038&YO-1m\\
8491.9157947&0.0004482&1056&TESS\\
8494.7259289&0.0005524&1057&TESS\\
8497.5365477&0.0004330&1058&TESS\\
8500.3478024&0.0004888&1059&TESS\\
8505.9685566&0.0004930&1061&TESS\\
8508.7796351&0.0004426&1062&TESS\\
8511.5901344&0.0004888&1063&TESS\\
8514.3994518&0.0004482&1064&TESS\\
8528.4546490&0.0009200&1069&ETD\\
8528.4557390&0.0009200&1069&ETD\\
8573.4253810&0.0009200&1085&ETD\\
8806.7005660&0.0011600&1168&ETD\\
8882.5887080&0.0011400&1195&ETD\\
8882.5893000&0.0009000&1195&\citet{RN786}\\
8885.4015610&$^{+0.0015510}_{-0.0014810}$&1196&\citet{RN699}\\
8885.4011690&$^{+0.0006960}_{-0.0007720}$&1196&\citet{RN699}\\
8902.2651870&$^{+0.0008220}_{-0.0009760}$&1202&\citet{RN699}\\
8919.1281410&$^{+0.0020980}_{-0.0016200}$&1208&\citet{RN699}\\
8933.1758110&$^{+0.0017660}_{-0.0017940}$&1213&\citet{RN699}\\
8944.4227790&0.0006700&1217&ETD\\
9191.7550600&0.0009700&1305&ETD\\
9194.5669500&0.0016300&1306&ETD\\
9231.1039217&0.0005250&1319&TESS\\
9233.9144487&0.0004940&1320&TESS\\
9236.7243179&0.0005188&1321&TESS\\
9236.7251900&0.0014100&1321&ETD\\
9239.5357656&0.0004714&1322&TESS\\
9245.1565895&0.0004347&1324&TESS\\
9247.9672925&0.0004558&1325&TESS\\
\hline\hline
\end{tabular}
\end{table}

\setcounter{table}{1}

\begin{table}
\caption{Continued.}             
\label{tab2}      
\centering  
\begin{tabular}{c c c c}  
\hline\hline       
Mid-time & Error & Cycle number & Source \\
($\mathrm{BJD_{TDB}-2450000}$) & (days) & &  \\
\hline
9250.7776903&0.0004371&1326&TESS\\
9250.7751800&0.0012100&1326&ETD\\
9253.5871352&0.0006276&1327&TESS\\
9253.5907500&0.0008000&1327&ETD\\
9256.3947700&0.0012000&1328&ETD\\
\hline\hline
\end{tabular}
\end{table}

\begin{table}
\caption{System parameters of WASP-35.}             
\label{tab3}      
\centering          
\begin{tabular}{c c c c}    
\hline\hline       
Parameter & \citet{RN516} & This work \\ 
\hline
Orbital period (days) & 3.161575$\pm$0.000002 & 3.1615691$\pm$0.0000003\\
Transit epoch (BJD-2450000) & 5531.47984$\pm$0.00015 & 5531.47920$\pm$0.00029\\
Transit Duration (days) & 0.1278$\pm$0.0009 & 0.1304$\pm$0.0004\\
Planet/star area ratio & 0.0154$\pm$0.0002 &0.0153$\pm$0.0001\\
Impact parameter & 0.30$^{+0.06}_{-0.09}$ &0.299$\pm$0.003 \\
Stellar reflex velocity (m/s) & 94.82$^{+7.11}_{-7.18}$ &98.90$\pm$3.64 \\
Centre-of-mass velocity (km/s)&17.718$\pm$0.004&17.7284$\pm$0.0003 \\
Orbital separation (AU)&0.04317$\pm$0.00033&0.04360$\pm$0.00020 \\
Orbital inclination (deg)&87.96$^{+0.62}_{-0.49}$&87.95$\pm$0.33 \\
Orbital eccentricity& 0 (fixed) & 0 (fixed)\\
Stellar mass ($M_{\odot}$)&1.07$\pm$0.03&1.106$\pm$0.015 \\
Stellar radius ($R_{\odot}$)&1.09$\pm$0.03&1.122$\pm$0.016 \\
Stellar density ($\rho_{\odot}$)&0.83$\pm$0.07&0.784$\pm$0.031\\
Stellar surface gravity (cgs)&4.40$\pm$0.02&4.381$\pm$0.011 \\
Stellar metallicity &-0.15$\pm$0.09&-0.051$\pm$0.051 \\
Stellar effective temperature (K) &5990$\pm$90&6072$\pm$63 \\
Planet mass ($M_{Jup}$)&0.72$\pm$0.06&0.765$\pm$0.029 \\
Planet radius ($R_{Jup}$)&1.32$\pm$0.05&1.349$\pm$0.022 \\
Planet density ($\rho_{Jup}$)&0.32$\pm$0.04&0.311$\pm$0.019 \\
Planet surface gravity (log $g_{Jup}$)&2.98$\pm$0.04&2.983$\pm$0.021 \\
Planet effective temperature ($A$=0) (K) &1450$\pm$30&1484$\pm$18 \\
\hline                  
\end{tabular}
\end{table}

\begin{table}
\caption{System parameters of HAT-P-30.}             
\label{tab4}      
\centering          
\begin{tabular}{c c c c}    
\hline\hline       
Parameter & \citet{RN518} & \citet{RN516} & This work \\ 
\hline
Orbital period (days) & 2.810595$\pm$0.000005 & 2.810603$\pm$0.000008& 2.8106006$\pm$0.0000004\\
Transit epoch (BJD-2450000) & 5456.46561$\pm$0.00037 & 5571.70135$\pm$0.00016&5523.92157$\pm$0.00043\\
Transit Duration (days) & 0.0887$\pm$0.0015 & 0.0920$\pm$0.0008 & 0.0916$\pm$0.0004\\
Planet/star area ratio & 0.0128$\pm$0.0004 &0.0122$\pm$0.0002& 0.0124$\pm$0.0001\\
Impact parameter & 0.854$^{+0.008}_{-0.010}$ &0.87$\pm$0.01 & 0.872$\pm$0.003\\
Stellar reflex velocity (m/s) & 92.67$\pm$2.50 &97.70$^{+6.19}_{-6.40}$ & 92.67$\pm$2.50\\
Centre-of-mass velocity (km/s)&45.51$\pm$0.18&44.677$\pm$0.001 & 44.6708$\pm$0.0002\\
Orbital separation (AU)&0.0419$\pm$0.0005&0.04118$\pm$0.00031 & 0.04114$\pm$0.00030\\
Orbital inclination (deg)&83.6$\pm$0.4&82.48$^{+0.16}_{-0.15}$ & 82.56$\pm$0.08\\
Orbital eccentricity&0.035$\pm$0.024& 0 (fixed) & 0 (fixed)\\
Stellar mass ($M_{\odot}$)&1.242$\pm$0.041&1.18$\pm$0.03 & 1.175$\pm$0.025\\
Stellar radius ($R_{\odot}$)&1.215$\pm$0.051&1.33$\pm$0.03 & 1.314$\pm$0.015\\
Stellar density ($\rho_{\odot}$)&&0.50$\pm$0.02& 0.517$\pm$0.012\\
Stellar surface gravity (cgs)&4.36$\pm$0.03&4.26$\pm$0.01 & 4.270$\pm$0.007\\
Stellar metallicity &+0.13$\pm$0.08&-0.08$\pm$0.08 &-0.079$\pm$0.079\\
Stellar effective temperature (K) &6304$\pm$88&6250$\pm$100 &6252$\pm$100\\
Planet mass ($M_{Jup}$)&0.711$\pm$0.028&0.76$\pm$0.05 &0.723$\pm$0.023\\
Planet radius ($R_{Jup}$)&1.340$\pm$0.065&1.42$\pm$0.03 &1.426$\pm$0.020\\
Planet density ($\rho_{Jup}$)&0.28$\pm$0.04&0.26$\pm$0.03 &0.249$\pm$0.011\\
Planet surface gravity (log $g_{Jup}$)&2.99$\pm$0.04&2.93$\pm$0.03 &2.910$\pm$0.015\\
Planet effective temperature ($A$=0) (K) &1630$\pm$42&1710$\pm$30 &1704$\pm$28\\
\hline                  
\end{tabular}
\end{table}

\begin{figure}
   \centering
   \includegraphics[width=\hsize]{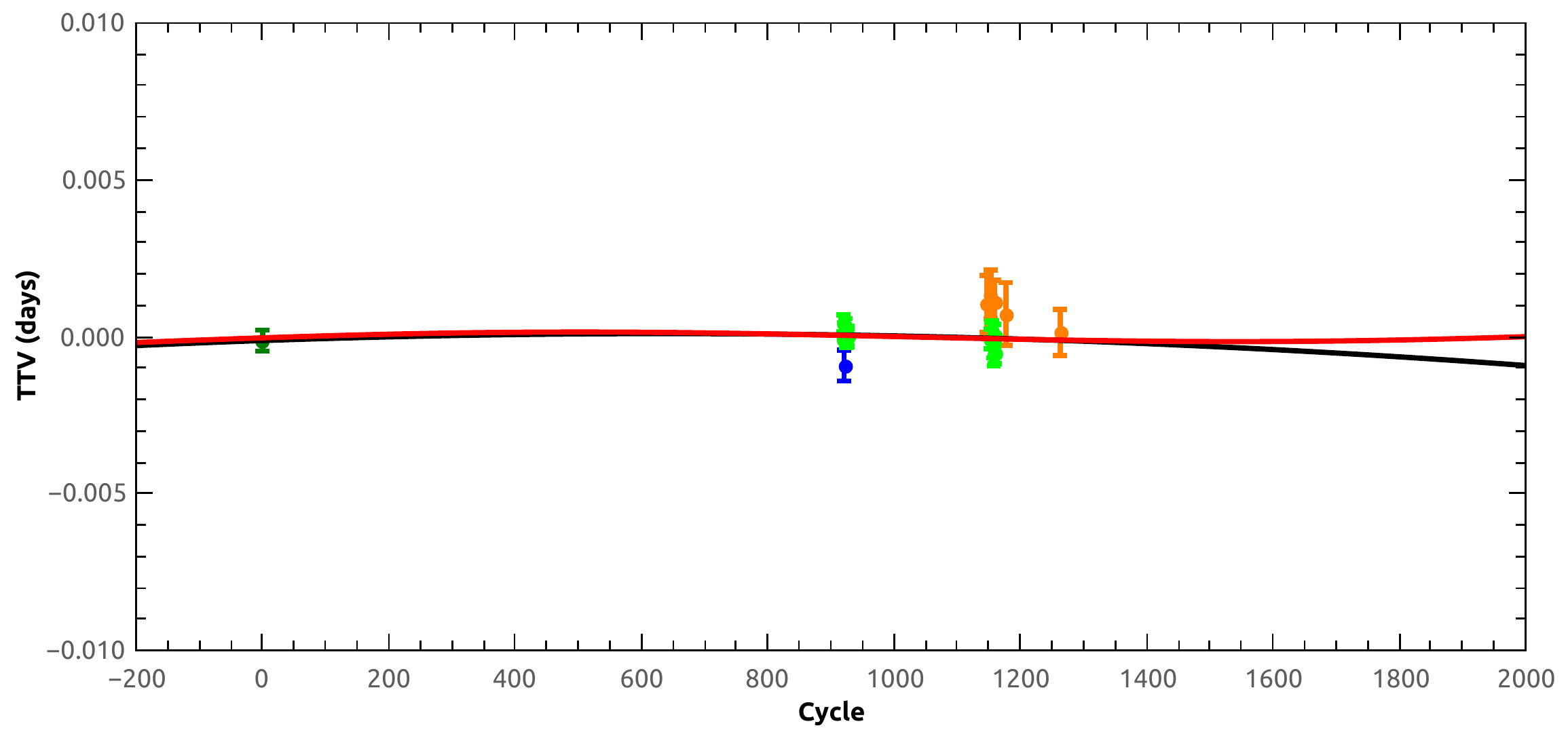}
 \caption{Transit timing variations after subtracting the constant-period model of WASP-35. The green and blue dots are the new transit times from TESS and YO-1m, the olive ones are from \citet{RN516}, and the orange ones denote those data points from ETD. The black curve shows the expected orbital decay model and the red curve shows the apsidal precession model.}
         \label{fig7}
\end{figure}

\begin{figure}
   \centering
   \includegraphics[width=\hsize]{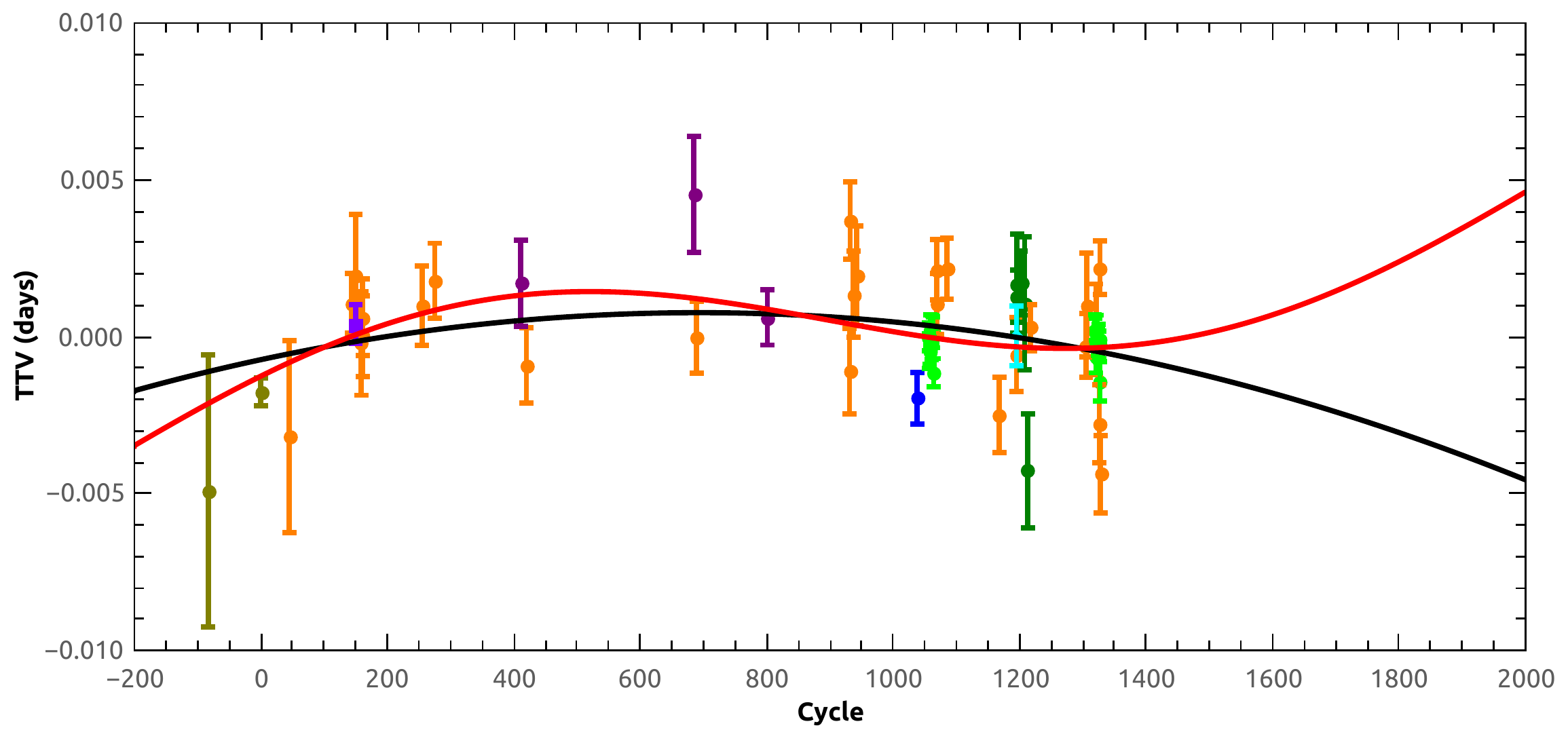}
      \caption{Transit timing variations after subtracting the constant-period model of HAT-P-30. The green and blue dots are the new transit times from TESS and YO-1m, the dark yellow ones are refined transit times from the data of \citet{RN518}, the violet one is from the data of \citet{RN517}, the purple ones are from \citet{RN700}, the olive ones are from \citet{RN699}, the cyan one is from \citet{RN786} and the orange ones denote those from ETD. The black curve shows the expected orbital decay model and the red curve shows the apsidal precession model.}
         \label{fig8}
\end{figure}

\begin{figure}
   \centering
   \includegraphics[width=\hsize]{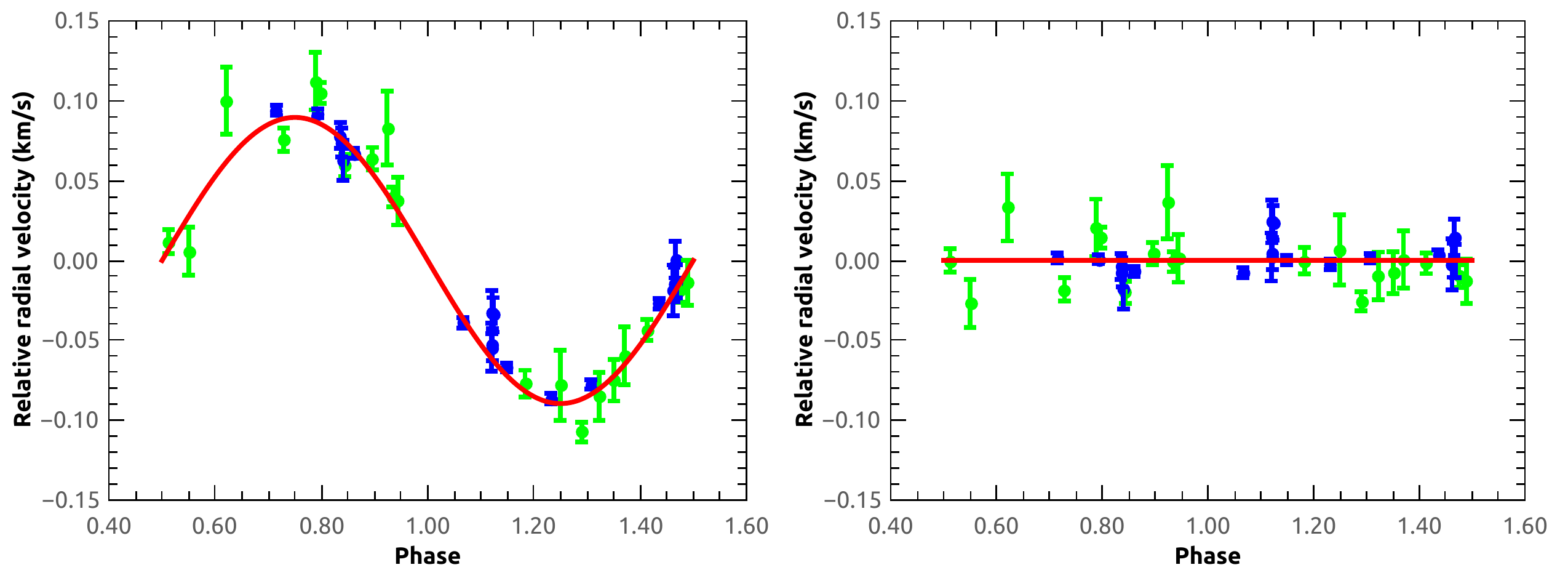}
      \caption{The RV data of HAT-P-30 with the linear trend best-fitting model and the residuals (right panel). The green dots are the RV data from \citet{RN516} and the blue ones are from \citet{RN518}.}
         \label{fig9}
   \end{figure}

\begin{figure}[!htbp]
    \centering
    \subfigure[]{
      \includegraphics[width=0.31\textwidth]{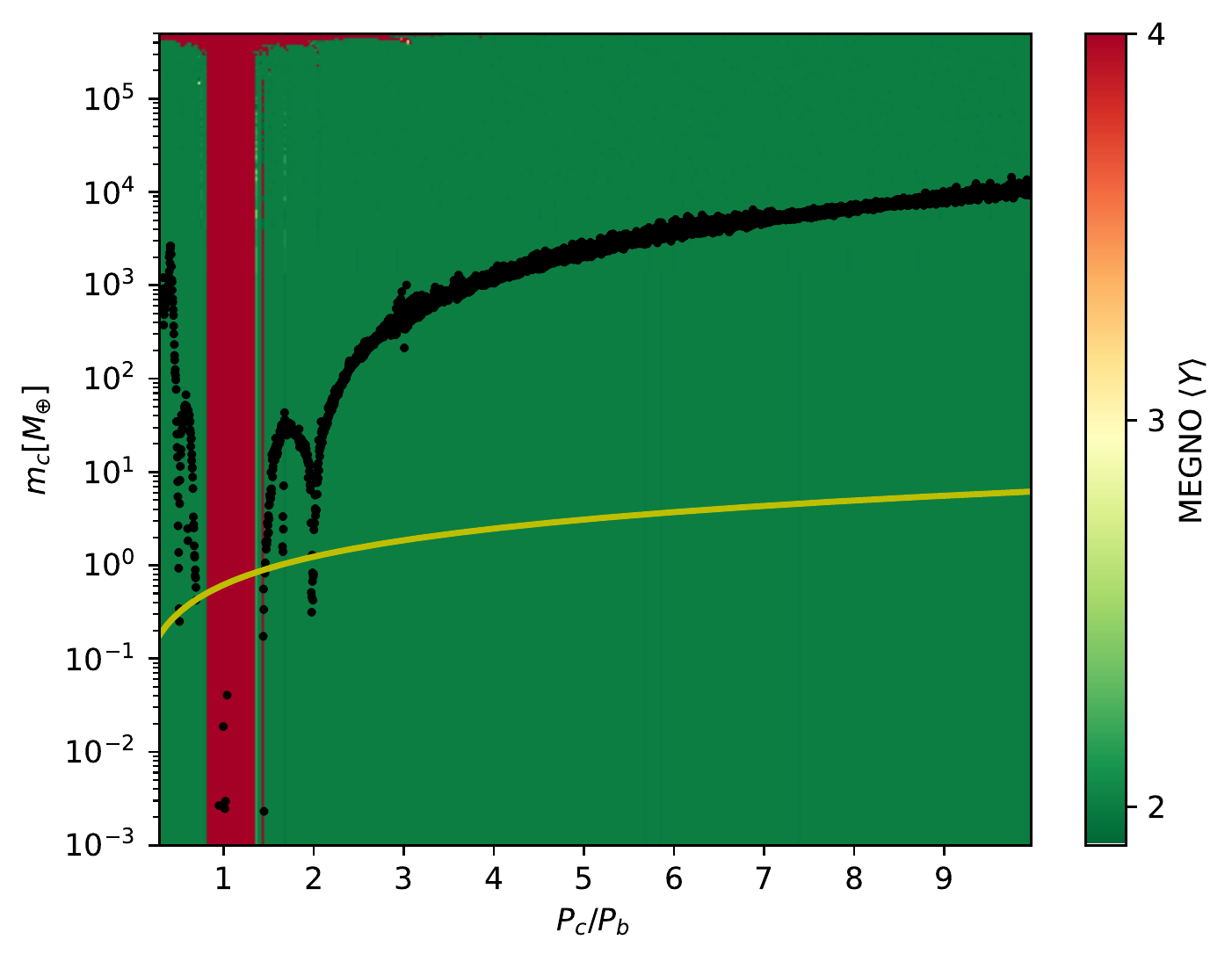}
      \label{fig10a}
      }
    \subfigure[]{
      \includegraphics[width=0.31\textwidth]{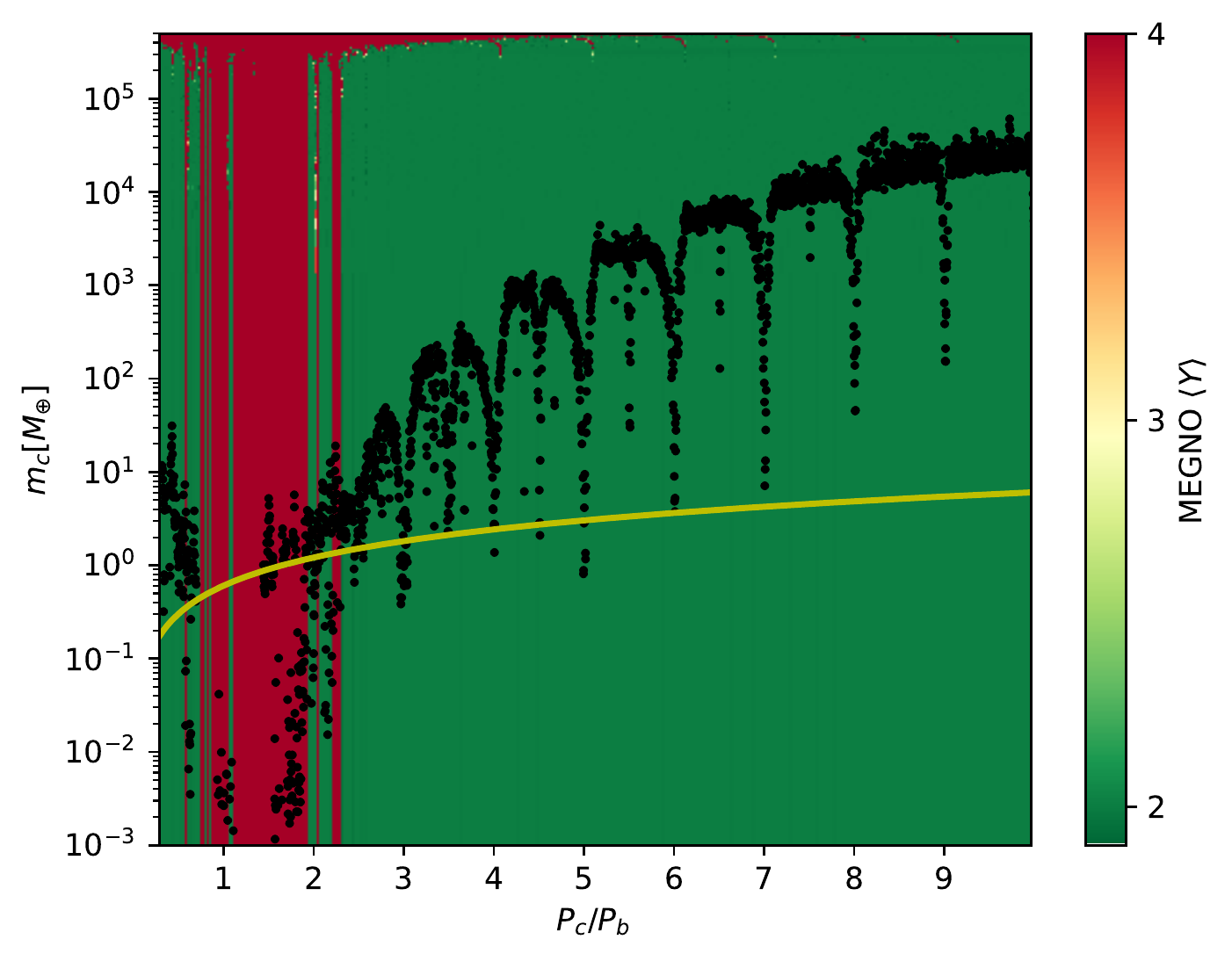}
      \label{fig10b}
      }
      \subfigure[]{
      \includegraphics[width=0.31\textwidth]{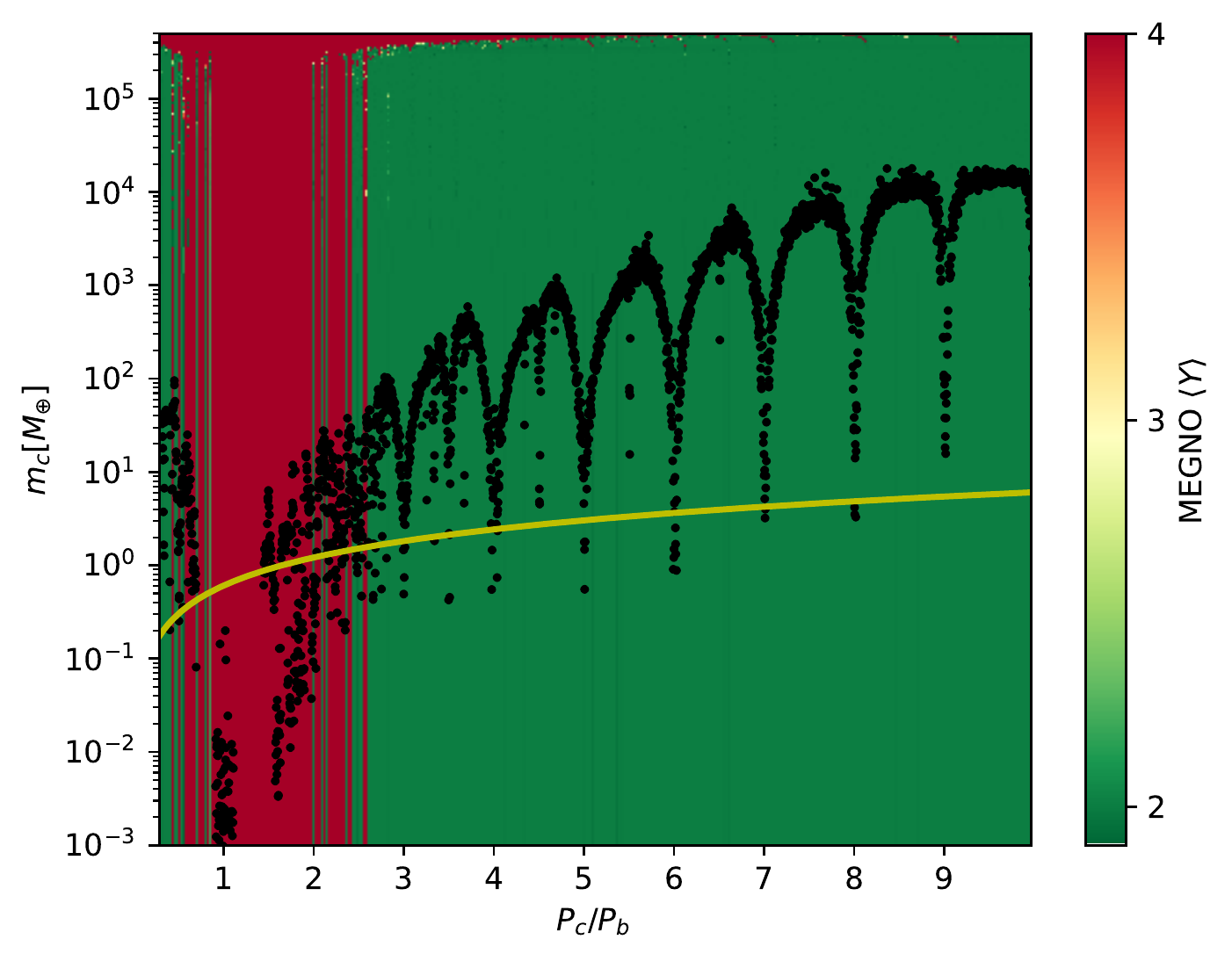}
      \label{fig10c}
      }
    \caption{The MEGNO maps of WASP-35 with the perturber initially on coplanar + circular orbit (a), coplanar+slightly eccentric orbit (e = 0.2) (b), inclined+slightly eccentric orbit (e = 0.2) (c), respectively. The dark dots represent the upper mass limit of the RMS of WASP-35b's TTVs, while the yellow curves are from the constraints of RMS of WASP-35's RV residuals after removing the components of WASP-35b. The regular orbital configurations denoted with green are distinguished from the chaotic ones labeled with red on the MEGNO map.}
    \label{fig10}
\end{figure}

\begin{figure}[!htbp]
    \centering
    \subfigure[]{
      \includegraphics[width=0.31\textwidth]{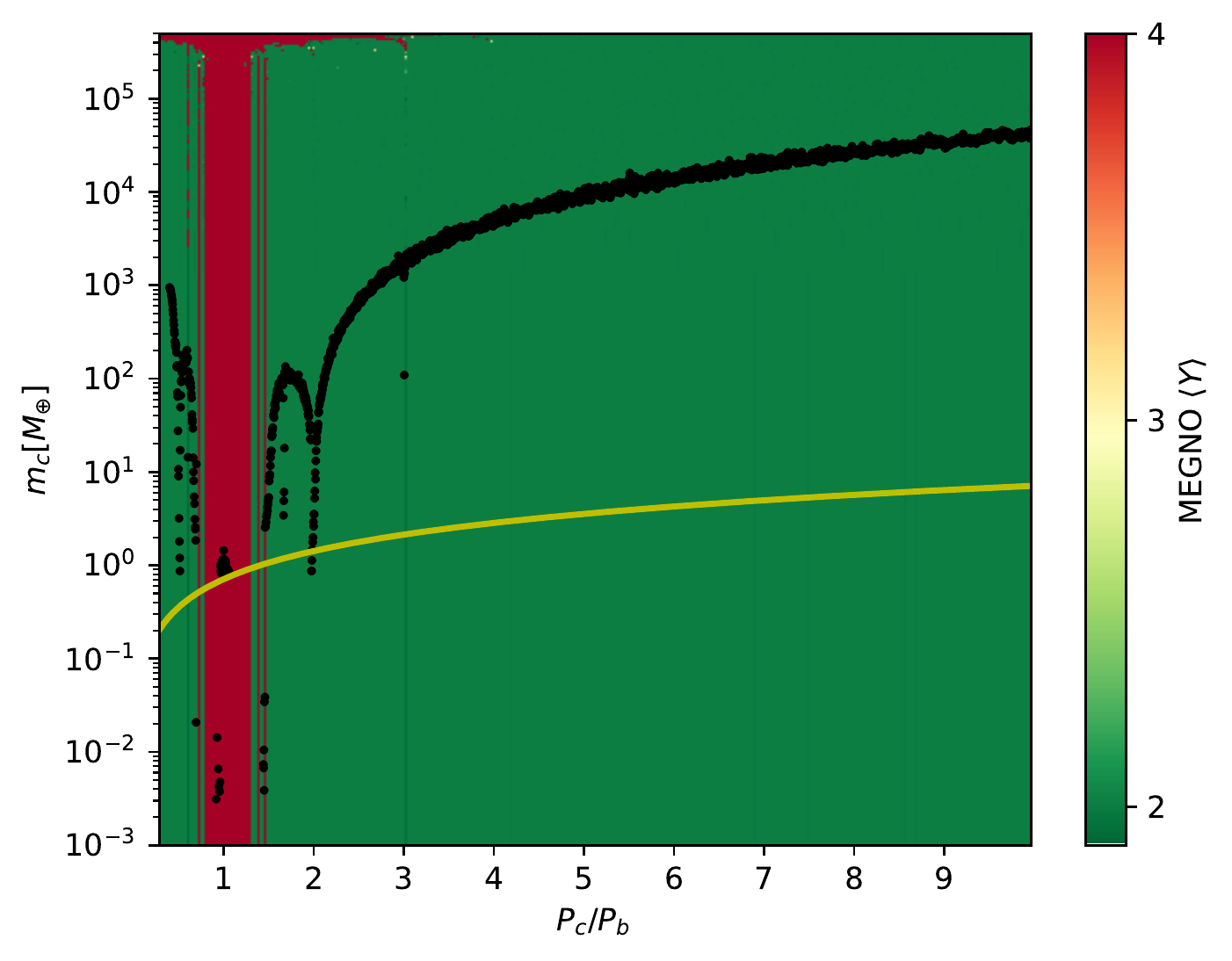}
      \label{fig11a}
      }
    \subfigure[]{
      \includegraphics[width=0.31\textwidth]{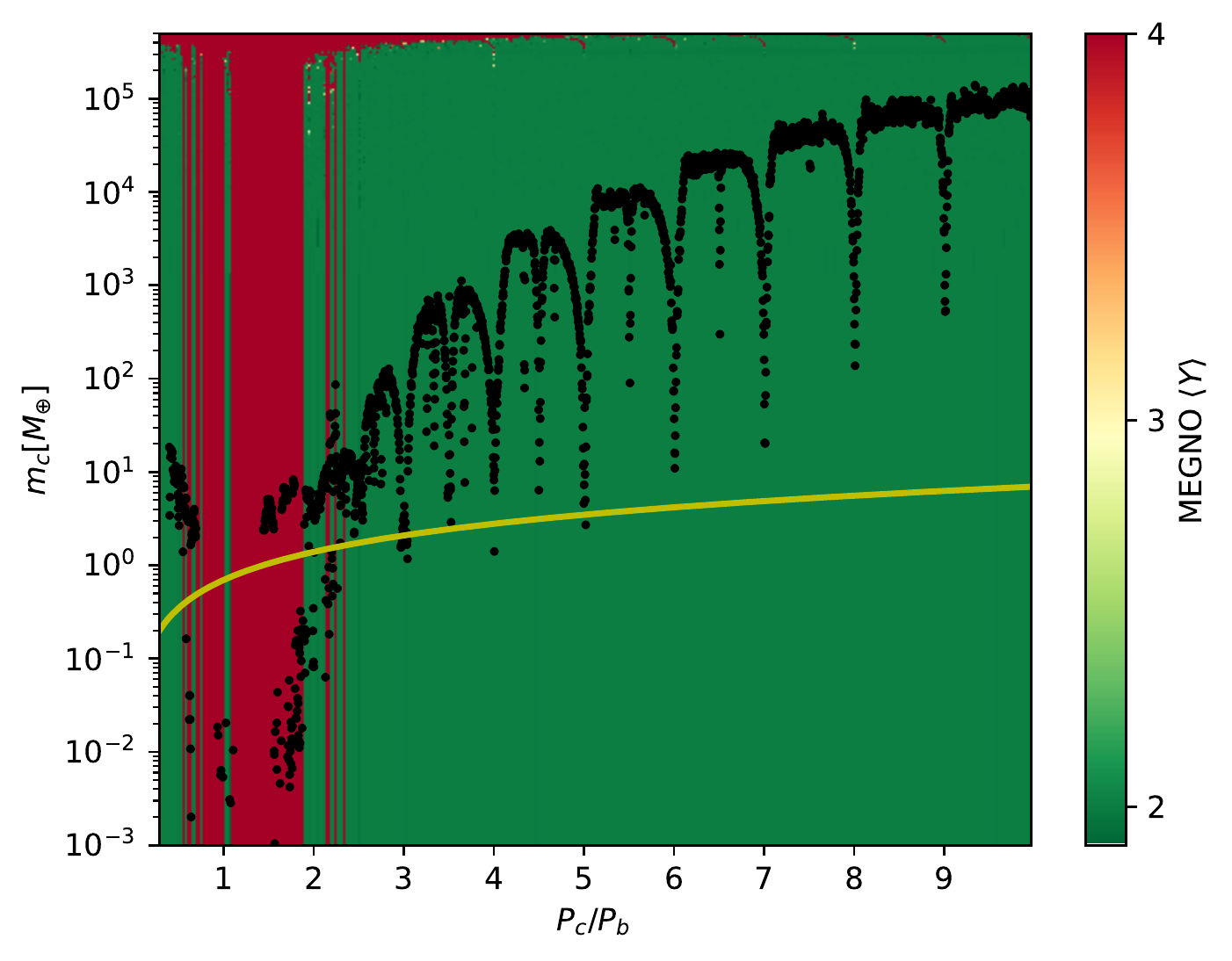}
      \label{fig11b}
      }
      \subfigure[]{
      \includegraphics[width=0.31\textwidth]{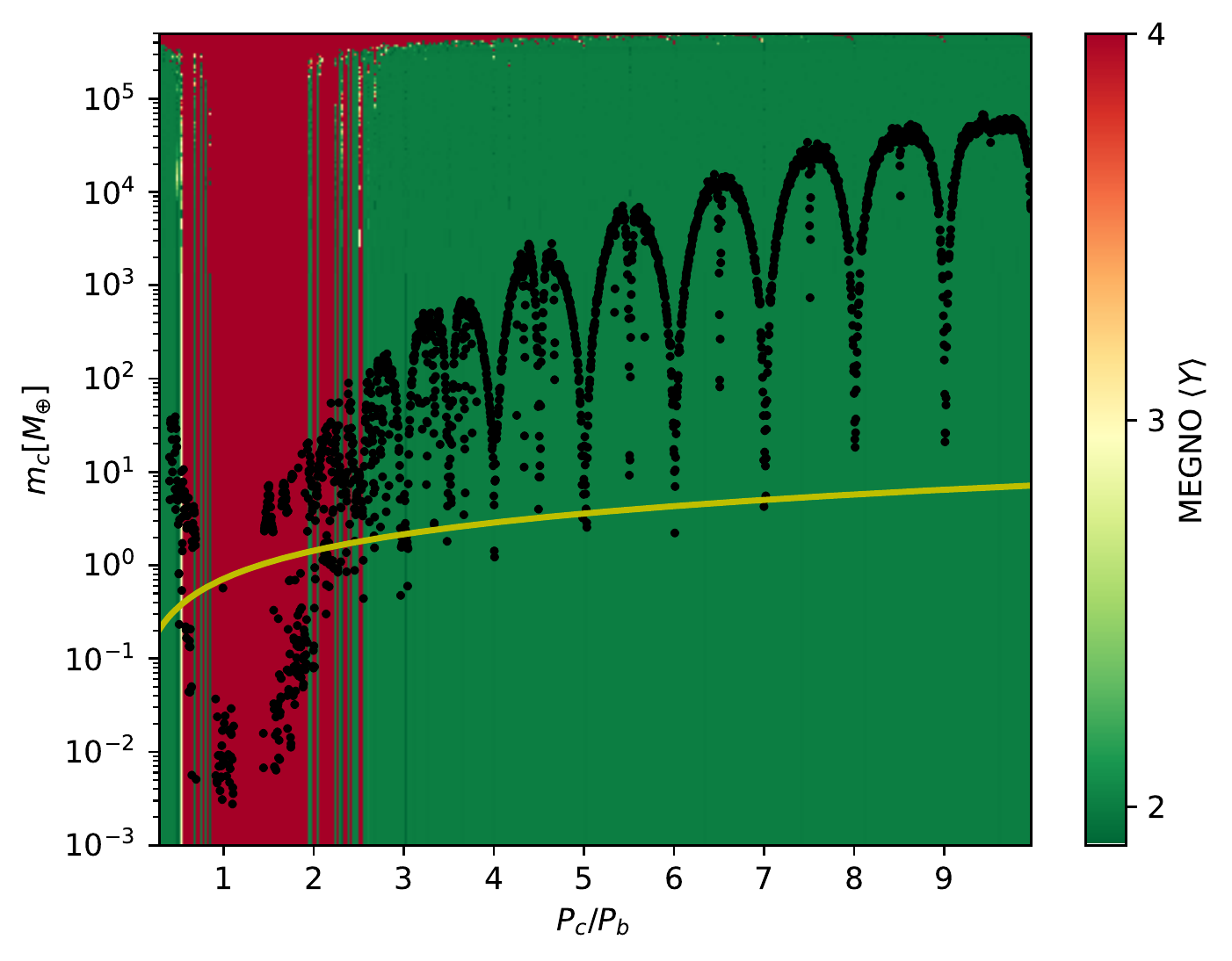}
      \label{fig11c}
      }
    \caption{The MEGNO maps of HAT-P-30 with the perturber initially on coplanar + circular orbit (a), coplanar+slightly eccentric orbit (e = 0.2) (b), inclined+slightly eccentric orbit (e = 0.2) (c), respectively. The dark dots represent the upper mass limit of the RMS of HAT-P-30b's TTVs, while the yellow curves are from the constraints of RMS of HAT-P-30's RV residuals after removing the components of HAT-P-30b. The regular orbital configurations denoted with green are distinguished from the chaotic ones labeled with red on the MEGNO map.}
    \label{fig11}
\end{figure}

\section{TTV modeling and prospect on atmospheric properties}\label{sec:dis}

\subsection{TTV signals} \label{ttvsection}

Transit timing variations are generally parameterized with the deviations between the observed transit times and their expected values assuming a Keplerian motion for the planet. The TTVs for these two targets were computed based on our new observations, TESS photometry, transit observations collected from ETD as well as transit observations in the literature. In tables \ref{tab1} and \ref{tab2}, there exist discrepancies between the transit times derived from simultaneous TESS and ground-based observations for some transit events, which may mean that the timing uncertainties were slightly underestimated, especially for the ground-based transit light curves which have lower quality than the TESS ones. Thus, we added a timing jitter to cope with this issue in the following TTV modeling. Refined orbital periods were obtained by linearly fitting all available transit times, in terms of a total of 55 transit times of HAT-P-30b, and 22 of WASP-35b. Figure \ref{fig7} presents all the TTV measurements for transiting exoplanet WASP-35b, and Figure \ref{fig8} for HAT-P-30b.

As the degeneracy in the explanation for hot Jupiter's TTVs mentioned in Section \ref{sec:intro}, we used four different models, namely linear ephemeris model, orbital decay model, the elliptical orbit precession model, and planetary interaction-induced TTV model, to fit the timing data of HAT-P-30b and WASP-35b. In this section we describle our analyses on the first three models, while the last one will be presented in Section \ref{addper}. Our timing analysis was partly similar to \citet{RN212}, \citet{RN209} and \citet{RN214}. 

The first model assumes a constant period:
\begin{displaymath}
		T_{tra}(E) = T_{0} + P \times E 
\end{displaymath}
where $T_{0}$ is the reference mid-transit time and $E$ is the cycle number.

The second model assumes the planet has a constant orbital period decay rate:
\begin{displaymath}
		T_{tra}(E) = T_{0} + P \times E + \frac{1}{2}\frac{dP}{dE}E^2
\end{displaymath}
where $dP/dE$ is the decay rate.

The third model assumes the planet has a nonzero orbital eccentricity $e$ and the argument of pericenter $\omega$ is precessing uniformly over time:
\begin{displaymath}
		T_{tra}(E) = T_{0} + P_{s} \times E - \frac{eP_{a}}{\pi}\mathrm{cos} \omega(E)
\end{displaymath}
\begin{displaymath}
		\omega(E) = \omega_{0} +  \frac{d\omega}{dE}E
\end{displaymath}
\begin{displaymath}
		P_{s} = P_{a}\bigg ( 1-\frac{1}{2\pi}\frac{d\omega}{dE}\bigg )
\end{displaymath}
where $P_{s}$ is the sidereal period, $P_{a}$ is the anomalistic period and $d\omega/dE$ is the precession rate \citep{RN265}.

For the these models, we preformed the MCMC calculations to find the best-fitting parameters using the \textbf{emcee} package \citep{RN263} and ran 50,000 MCMC steps with 1,000 burn-in steps to ensure convergence. The results of three timing model fitting are listed in Table\ref{tab5} and Table \ref{tab6}, and Figure \ref{fig7} and Figure \ref{fig8} show the transit timing data with the best-fitting orbital decay and apsidal precession models.

We used the Bayesian Information Criterion \citep[BIC;][]{RN267} as the penalty to compare two different best-fit models, and the BIC is defined as:
\begin{displaymath}
		\mathrm{BIC} = \chi^{2} + k \mathrm{log} n
\end{displaymath}
where $k$ is the number of free parameters and $n$ is the number of data points.

\subsubsection{WASP-35}

Figure \ref{fig7} is the linear plot of TTV versus cycle for this planet. The deviations of the transit times from the linear ephemeris have a RMS of 37 s. Compared with the constant period model ($\chi^{2}_{min}=22.807$) and the apsidal precession model ($\chi^{2}_{min}=22.254$), the orbital decay model had a lower minimum chi-squared ($\chi^{2}_{min}=22.168$). But we found that the constant period model had the lowest BIC value, that is, the constant period model was the favored model with $\Delta(\mathrm{BIC_{1,2}}) = 2.452$ and $\Delta(\mathrm{BIC_{1,3}}) = 8.720$. We assumed a multivariate Gaussian distribution for the posterior of all the parameters, the Bayes factor $B$:

\begin{displaymath}
		B_{1,2} = \mathrm{EXP} [-\Delta(\mathrm{BIC})/2] = 3.408
\end{displaymath}
\begin{displaymath}
		B_{1,3} = \mathrm{EXP} [-\Delta(\mathrm{BIC})/2] = 78.257
\end{displaymath}
This means that the observations of WASP-35 slightly favored the constant period model.

\begin{table}
\caption{Timing model parameters of WASP-35.}             
\label{tab5}      
\centering                         
\begin{tabular}{c c c c}        
\hline\hline                 
Parameter & Symbol & Unit & Value \\   
\hline                        
\hline
\multicolumn{4}{l}{\textbf{Constant period model}} \\
Period & $P$ & days & $3.1615691\pm0.0000003$\\
Mid-transit time & $T_{0}$ & $\mathrm{BJD_{TDB}}$ & $2455531.47920\pm0.00029$ \\
\hline
$N_{dof}$ & & & 20 \\
$\chi^{2}_{min}$ & & & 22.807 \\
BIC & & & 28.989 \\
\hline      
\hline
\multicolumn{4}{l}{\textbf{Orbital decay model}} \\    
Period & $P$ & days & $3.1615698\pm0.0000009$ \\
Mid-transit time & $T_{0}$ & $\mathrm{BJD_{TDB}}$ & $2455531.47907\pm0.00033$  \\
Decay rate & $dP/dE$ & days/orbit & $-1.09\times10^{-9}\pm1.36\times10^{-9}$ \\
\hline
$N_{dof}$ & & & 19\\
$\chi^{2}_{min}$ & & & 22.168\\
BIC & & & 31.441\\
\hline       
\hline     
\multicolumn{4}{l}{\textbf{Apsidal precession model}} \\    
Sidereal period & $P_{s}$ & days & $3.1615693\pm0.0000004$ \\
Mid-transit time & $T_{0}$ & $\mathrm{BJD_{TDB}}$ & $2455531.47902\pm0.00033$\\
Orbital eccentricity & e & & $0.00024\pm0.00017$\\
Argument of periastron & $\omega_{0}$ & rad & $2.145\pm0.540$ \\
Precession rate & $d\omega/dE$ & rad/orbit & $0.00244\pm0.00170$ \\
\hline
$N_{dof}$ & & & 17\\
$\chi^{2}_{min}$ & & & 22.254\\
BIC & & & 37.709\\
\hline
\hline       
\end{tabular}
\end{table}

\begin{table}
\caption{Timing model parameters of HAT-P-30.}             
\label{tab6}      
\centering                        
\begin{tabular}{c c c c}        
\hline\hline                 
Parameter & Symbol & Unit & Value \\   
\hline                        
\hline
\multicolumn{4}{l}{\textbf{Constant period model}} \\
Period & $P$ & days & $2.8106006\pm0.0000004$\\
Mid-transit time & $T_{0}$ & $\mathrm{BJD_{TDB}}$ & $2455523.92157\pm0.00043$ \\
\hline
$N_{dof}$ & & & 53 \\
$\chi^{2}_{min}$ & & & 119.220 \\
BIC & & & 127.235 \\
\hline      
\hline
\multicolumn{4}{l}{\textbf{Orbital decay model}} \\    
Period & $P$ & days & $2.8106049\pm0.0000019$ \\
Mid-transit time & $T_{0}$ & $\mathrm{BJD_{TDB}}$ & $2455523.92083\pm0.00047$  \\
Decay rate & $dP/dE$ & days/orbit & $-6.24\times10^{-9}\pm2.77\times10^{-9}$ \\
\hline
$N_{dof}$ & & & 52\\
$\chi^{2}_{min}$ & & & 113.793\\
BIC & & & 125.815\\
\hline       
\hline     
\multicolumn{4}{l}{\textbf{Apsidal precession model}} \\    
Sidereal period & $P_{s}$ & days & $2.8106047\pm0.0000011$ \\
Mid-transit time & $T_{0}$ & $\mathrm{BJD_{TDB}}$ & $2455523.91840\pm0.00090$\\
Orbital eccentricity & e & & $0.00323\pm0.00065$\\
Argument of periastron & $\omega_{0}$ & rad & $2.290\pm0.325$ \\
Precession rate & $d\omega/dE$ & rad/orbit & $0.00270\pm0.00050$ \\
\hline
$N_{dof}$ & & & 50\\
$\chi^{2}_{min}$ & & & 100.051\\
BIC & & & 120.088\\
\hline
\hline       
\end{tabular}
\end{table}

\subsubsection{HAT-P-30}

Figure \ref{fig8} is the linear plot of TTV versus cycle for this planet. Compared with the TTV analysis of \citet{RN784}, our data set covers a longer time baseline and contains more data points, which exhibits a secular timing effect of HAT-P-30b. The deviations of the transit times from the linear ephemeris have a RMS of 158 s. Compared with the constant period model ($\chi^{2}_{min}=119.220$) and the orbital decay model ($\chi^{2}_{min}=113.793$), the apsidal precession model had a lower minimum chi-squared ($\chi^{2}_{min}=100.051$). And we also found that the apsidal precession model had the lowest BIC value, that is, the apsidal precession model was the favored model with \textbf{$\Delta(\mathrm{BIC_{3,1}}) = 19.744$} and $\Delta(\mathrm{BIC_{3,2}}) = 9.095$. We assumed a multivariate Gaussian distribution for the posterior of all the parameters, the Bayes factor $B$:

\begin{displaymath}
		B_{3,1} = \mathrm{EXP} [-\Delta(\mathrm{BIC})/2] = 35.641
\end{displaymath}
\begin{displaymath}
		B_{3,2} = \mathrm{EXP} [-\Delta(\mathrm{BIC})/2] = 17.523
\end{displaymath}

This means slightly favored to the apsidal precession model for the observations.

According to the constant-phase lag model for tidal evolution suggested by \citet{RN269}, the decay rate is defined as:
\begin{displaymath}
		\frac{dP}{dt} = -\frac{27\pi}{2Q^{'}_{*}}\bigg(\frac{M_{p}}{M_{*}}\bigg)\bigg(\frac{R_{*}}{a}\bigg)
\end{displaymath}
where $Q^{'}_{*}$ is the modified quality factor of the stellar tidal oscillations, $M_{p}$ is the planet mass, $M_{*}$ is the stellar mass. For the HAT-P-30, this yields
\begin{displaymath}
		\frac{dP}{dt} = -553.08\pm245.52~\mathrm{ms~yr^{-1}}
\end{displaymath}
Based on the decay rate of the orbital decay model, we derived the modified quality factor of 

\begin{displaymath}
		Q^{'}_{*}=(8.10\pm3.62)\times10^{2}
\end{displaymath}

This value is significantly lower compared with the typical values of $10^{5}-10^{7}$ for binary star systems \citep{RN273,RN271} and $10^{5}-10^{6.5}$ for hot Jupiters \citep{RN275,RN277,RN279}. 

In addition, we also considered whether this TTV is caused by the R\o mer effect like WASP-4 system as \citep{RN787,RN788}. Because of the Doppler effect, if there is any line-of-sight acceleration of the system, it would lead to a decay of the orbital period:
\begin{displaymath}
		\frac{dP}{dt} = \frac{\dot{v}_{r}P}{c}
\end{displaymath}
where $\dot{v}_{r}$ represents the line-of-sight acceleration of the radial motion. We performed an independent RV modeling to test this possibility. We modeled all of the RV measurements simultaneously and other configurations followed the same procedure as above, except for a long-term linear trend $\dot{v}_{r}$. Compared with the RV model with a long-term linear trend ($\chi^{2}_{2}=46.291$, $\mathrm{BIC_{2}}=52.610$, see Figure \ref{fig9}), the RV model without a long-term linear trend has a lower $\chi^{2}$ ($\chi^{2}_{1}=46.260$) and BIC value ($\mathrm{BIC_{1}}=50.999$), namely the best-fitting result prefers the RV model without a long-term linear trend. We assumed a multivariate Gaussian distribution for the posterior of all the parameters, the Bayes factor $B$: 
\begin{displaymath}
		B_{1,2} = \mathrm{EXP} [-\Delta(\mathrm{BIC})/2] = 2.237
\end{displaymath}
\textbf{This means the difference between two RV models is not more than a bare mention.}

We preformed the MCMC calculations to derive the value of line-of-sight acceleration $\dot{v}_{r}$. The configurations of MCMC calculations are as above, and the value of line-of-sight acceleration 
\begin{displaymath}
		\dot{v}_{r} = -0.013\pm0.017~\mathrm{m~s^{-1}~day^{-1}}
\end{displaymath}
For the HAT-P-30, this yields
 \begin{displaymath}
		\frac{dP}{dt} = -3.85\pm5.03~\mathrm{ms~yr^{-1}}
\end{displaymath}
Based on the line-of-sight acceleration we obtained, the implied period derivative is about two orders of magnitude smaller than the decay rate observed from transit timing, so the R\o mer effect cannot account for all of the observed period decrease.

\subsection{Upper Mass Limit of a Hypothetical Perturber}\label{addper}

The results from our TTV measurements (see Section~\ref{ttvsection}) allow to infer the upper mass limit for an additional perturbing planet in each of these two systems, which assuming that all the TTV singals origin from the gravitational interactions among planets involved. Although inverting the high-cadence and high signal-to-nosie TTV signals could be used to measure the mass and eccentricity of the perturbing exoplanets, obviously that is not our cases for HAT-P-30b and WASP-35b. However, the perturbation from the additional planet can be approximately quantified by the RMS of TTVs for our sparse measurements. The TTV effects are strongly amplified for orbital configurations in (or near) Mean-Motion Resonances (MMR) \citep{RN323,RN684}, in which the detection of a low-mass planetary perturbing body will in principle be permitted.

An upper mass limit can be obtained by employing a {\it N}-body code to perform direct orbit integrations, which is widely used in the literature \citep[e.g.,][]{RN734,RN733,RN735,RN700,RN749}. Within the framework of the three-body orbital configuration, we numerically integrated the orbits of each of these two hot Jupiters and a hypothetic perturbing planet around their host star. To meet our need, we modified the TTV inversion code of \citet{RN685}, which employs \textbf{TTVFast} \citep{RN663} to perform direct orbit integrations. Summarily, our new code caculates the RMS of synthetic TTVs, whose epochs are selected to be identical to those of measured TTVs, considering a transiting exoplanet perturbed by an $1 M_{\oplus}$ planet on an arbitrary orbital architecture. Such a configuration will help simplify the computing of upper mass limit, as the amplitude of a TTV pattern is linearly proportion to the mass of perturbing planet. Hence, the upper mass limit of perturbing planet could be well estimated by the product of the RMS of measured TTVs divided by that of synthetic TTVs.

This code was then applied to a series of orbital periods of the perturbing planet while fixing all the other orbital parameters. As the TTV patterns strongly depended on the perturber's mass, the orbital period, the eccentricity, and the mutual inclination of the orbit \citep{RN323,RN684,RN719,RN313,RN720}, we performed orbital integrations in terms of three different orbital architectures for hypothetical perturbers in this study: (a) initially on coplanar and circular orbit; (b) initially on coplanar and slightly eccentric orbit (i.e., $e_{c}=0.2$); (c) initially on inclined and slightly eccentric orbit (i.e., $i_{c}=i_{b}-30\degr$, $e_c=0.2$, where $i_b$ and $i_c$ denote the inclinations of the known transiting planet and the perturber, respectively). For clarity, hereafter we label these three different orbital architectures as Case a, Case b, and Case c, respectively. Except the orbital period of the perturbing planet, the remaining orbital elements (i.e., the longitude of the ascending node $\Omega$, the argument of periastron $\omega$, and the mean anomaly $M$ at a reference time) were fixed to selected values, $\Omega_{c}=\Omega_{b}=0$, $\omega_{c}=\omega_{b}=90\degr$, $M_{c}=M_{b}+180\degr$; the orbital period of the hypothetical perturber $P_{c}$ was searched from 1 day to 10$P_{b}$ with a step of 0.007 days. We expect that the first setting would provide a most conservative estimate of the upper mass limit of a possible perturber \citep{RN736,RN737,RN738,RN739,RN749}. 

In addition, the RVs of the host star induced by a hypothetical perturber could place constraints on the perturber's mass. If the planets are on non-interacting Keplerian orbits, the RVs of the host star are the sum of the RVs arosed by each planetary component's Keplerian motion \citep{RN717}. Therefore, the residuals of both HAT-P-30 and WASP-35's RVs after removing the contributions from both known planets could well present the RVs induced by additional bodies. Placing good constraints on the mass of the additional body through fitting RV curves requires a good coverage in its orbit phase and high signal-to-noise RV measurements, however, which is not the cases for the hypothetical perturbers in both HAT-P-30 and WASP-35. Therefore, we used the RMS of the residuals of both HAT-P-30 and WASP-35's RVs instead of the residuals to statistically constrain the mass of the hypothetical perturber. The amplitude of RV curves $K$ arosed by a planet on its host star is scaled with the following formula:
\begin{displaymath}
\left(\frac{M_p\sin i_p}{M_\oplus}\right)=11.19\left(\frac{K}{m/s}\right)\sqrt{1-e^2}\left(\frac{M_*}{M_\odot}\right)^{2/3}\left(\frac{P_{orb}}{1 yr}\right)^{1/3}
\end{displaymath}
where $M_p$, $M_*$, and $P_{orb}$ are the mass of the planet, the mass of the host star, and the orbital period of the planet, respectively. For the orbital eccentricties (i.e., $e_c\leq 0.2$) of the perturbers we assumed, the RMS of RV curves statistically equals about $\sqrt{2}/2$ times of the RV amplitude. We obtained the mass limits, that is, the yellow curves in Figure \ref{fig10} and Figure \ref{fig11} of the hypethetical perturbers based on the RMS of both HAT-P-30 and WASP-35's RV residuals derived in Section \ref{sec:tra}.

Some constraints could also be placed on the perturber's mass by the requirements of the long-term stability of perturber's orbits. For this purpose, we computed the Mean Exponential Growth factor of Nearby Orbits \citep[MEGNO;][]{RN740,RN741,RN742}, which was originally developed to study the global dynamics of non-axisymmetric galactic potentials, by employing \textbf{REBOUND} to perform direct orbital integrations and calculate associated variational equations of motion over a grid of initial values of orbital parameters \citep{RN668,RN743,RN744}. In addition to the orbital period of the perturber, here we also adjusted its mass; we integrated each initial grid point for 500 years (i.e., $\sim 10^4$ of orbital periods of transiting exoplanets), which will highlight the location of weak chaotic high-order mean-motion resonances. MEGNO is often used to quantitatively measure the degree of stochastic behaviour of a non-linear dynamical system and thus detect the chaotic resonances \citep{RN741,RN745}. In addition to integrating the Newtonian equations of motion, the associated variational equations of motion are caculated simultaneously for obtaining the MEGNO at each integration time step. Following \citet{RN740} and \citet{RN742}, the MEGNO index is defined as:
\begin{displaymath}
Y(t) = \frac{2}{T}\int_{0}^{T} \frac{||\dot{\delta}(t)||}{||\delta(t)||} t dt
\label{eq:megno}
\end{displaymath}
\noindent
where $\dot{\delta}/\delta$ is the relative change of the variational vector $\delta$. The time-averaged or mean $Y(t)$ is parameterized with:
\begin{displaymath}
\langle Y(t) \rangle = \frac{1}{T}\int_{0}^{T} Y(t) dt
\label{eq:megno_mean}
\end{displaymath}
\noindent
For the results in Figure~\ref{fig10} and Figure~\ref{fig11}, it is always the time-averaged MEGNO index that is utilized to quantitatively differentiate between quasi-periodic and chaotic dynamics. The regular orbits which evolve quasi-periodically in time $\langle Y \rangle$, will asymptotically approach 2.0 for $t \rightarrow \infty$, while for chaotic orbits it grows, proportionally to the Lyapunov exponent $\Lambda$, as ($\Lambda$/2)T. For a chaotic orbital evolution $\langle Y\rangle$ significantly deviates from 2.0 with orbital parameters exhibiting erratic temporal evolutions. Importantly, MEGNO is unable to prove whether a dynamical system is evolving quasi-periodically. This hints that a given system cannot be proven to be stable or bounded for all times. For a given initial condition once MEGNO detected chaotic behaviors, however, there is no doubt about its erratic nature in the future \citep{RN745}.

In the following parts, we present the results of each system for which we have calculated the RMS scatter of TTVs ($\rm TTV_{\rm RMS}$) on a series of orbital periods of a perturbing planet. In each of the three cases, we found the common instability regions located in the proximity of the transiting planet with MEGNO color-coded as yellow (corresponding to $\langle Y\rangle > 3.5$).

\subsubsection{WASP-35b}

Through over-plotting the $\rm TTV_{\rm RMS}$ for a certain value, we found similar results to those of \citet{RN323} and \citet{RN684}, which suggest that the TTVs are relatively more sensitive to orbital architectures involving MMRs. Furthermore, $\rm TTV_{\rm RMS}$ of both Case b and c are more complex, for which high-order MMRs (e.g., $P_c/P_b\simeq 4:1, 5:1$, and so on.) generate large TTV signals and thus the corresponding upper mass limit dramatically drops, comparing with that of Case a; interestingly, a perturbing body of only $\sim 0.606~M_{\oplus}$ located near 1:1 MMRs in Case b could well reproduce measured $\rm TTV_{\rm RMS}$ of WASP-35 b. While over-plotting the RMS of RV residuals ($\rm RV_{\rm RMS}$) for a certain value with removing the component due to known WASP-35b, we found that the constraints due to $\rm RV_{\rm RMS}$ on the upper mass limit are more stringent than those due to $\rm TTV_{\rm RMS}$. For HAT-P-30, similar conclusion has been drawn. As shown in Figure~\ref{fig10}, we found that a coplanar perturbing body of mass (upper limit) around $0.606-1~M_{\oplus}$ initially with circular orbit will cause a RMS of $37\,{\rm s}$ when located in the $P_c/P_b=$ 1:3 and 2:1. For the perturber on more inclined, and/or initially slightly eccentric orbits the high order resonances (i.e., $P_c/P_b=$ 4:1, 5:1 and 6:1 for Case b and even $P_c/P_b=$ 7:1 and 8:1 for Case c) appear beneath the upper mass limit of $\rm RV_{\rm RMS}$ in addition to the co-orbital configuration (i.e., 1:1 MMR).

\subsubsection{HAT-P-30b}
For the HAT-P-30 system the measured $\rm TTV_{\rm RMS}$ was $158 \rm s$. Additional bodies in Case a with an upper mass limit as low as $\sim 0.356~M_{\oplus}$ at the 1:2 and $\sim 1.423~M_{\oplus}$ at the 2:1 MMR could cause the observed TTV scatter. Hypothetical planets of $0.712~M_{\oplus}$, $1.423~M_{\oplus}$, $2.135~M_{\oplus}$, $2.846~M_{\oplus}$ , and $3.558~M_{\oplus}$ could well produce the observed $\rm TTV_{\rm RMS}$ located near 1:1, 2:1, 3:1, 4:1 and 5:1 exterior MMRs in Case b, respectively; and $2.135~M_{\oplus}$, $2.491~M_{\oplus}$, $2.846~M_{\oplus}$, $3.558~M_{\oplus}$, $4.270~M_{\oplus}$and $4.981~M_{\oplus}$ near 3:1, 7:2, 4:1, 5:1, 6:1 and 7:1 exterior MMRs in Case c, respectively. In both Case b and c, the structure of upper mass limits is quite complex from 2:1 to 3:1 exterior MMRs, see Figure~\ref{fig11} for further details.

\subsection{Prospect on atmospheric properties}

Transiting exoplanets offer the possibility of characterising the planetary  atmosphere using observations at different orbital phases. These observations include the transmission spectrum during the transit \citep{RN548,RN562}, the emission spectrum at secondary eclipse \citep{RN686,RN687} and the phase curve throughout the orbit\citep{RN688,RN689}. The transmission spectrum probes the wavelength-dependent extinction due to the planetary atmosphere at its day–night terminator region by using photometric or spectroscopic observations \citep{RN242,RN243,RN244}. For hot Jupiters, they have relatively low densities, large radius and hot equilibrium temperatures, which will lead to large atmospheric scale heights, namely, the planets have significant atmospheric structures and thus are ideal targets for atmospheric studies. Both WASP-35b and HAT-P-30b belong to hot Jupiters. Therefore, based on the new system parameters, we attempt to predict the atmospheric properties of WASP-35b and HAT-P-30b. 

For the transmission spectra, the amplitude of the transmission signals in a cloud-free atmosphere is proportional to the atmospheric scale height $H$:
\begin{displaymath}
		H= \frac{k_{B}T_{eq}}{{\mu}_{m}g_{b}}
\end{displaymath}
where $k_B$ is Boltzmann's constant, $T_{eq}$ is the temperature of the planet, $\mu_{m}$ is the mean molecular mass and $g_b$ is the surface gravity of the planet \citep{RN242,RN243,RN244}. According to \citet{RN423}, hot Jupiters should be H/He-dominated atmosphere, and the mean molecular mass is approximately 2.3 amu, so the atmospheric scale height for WASP-35b is about to 550 km and 750 km for HAT-P-30b. So, both WASP-35b and HAT-P-30b should have distinct atmospheres, the spectral features in transmission spectra could have amplitude of 5-10$H$ which are suitable for studying the properties of their atmospheres \citep{RN354}. 

For hot Jupiters, the spectral features in the optical transmission spectrum are expected to have Na, K, TiO and VO. According to the dichotomy of highly irradiated, close-in giant planets proposed by \citet{RN374} and the newly derived system parameters, both WASP-35b and HAT-P-30b are at the boundary between the pL and pM classes. Therefore we could expect that alkali metal absorption lines as well as TIO and VO likely appear in their high-precision transmission spectra. \citet{RN538} found the abundances of TiO and VO were strongly affected by the C/O ratio. In general, the upper atmospheric absorptions in the optical may lead to thermal inversions, such as the strong optical absorption of TiO and VO \citep{RN374} as well as the absorption of alkali metal at high C/O ratios \citep{RN430}. H$_{2}$O, CO, CH$_{4}$, CO$_{2}$ and H$_{2}$ are the major chemical and spectroscopic species which dominant the C/O ratio in the hot Jupiter atmosphere and the series of reactions can be summarized as 
\begin{displaymath}
		\mathrm{CO+3H_{2} \rightleftharpoons CH_{4}+H_{2}O}.
\end{displaymath}
WASP-35b and HAT-P-30b have relative high temperatures ($T_{eq}>1400$K), a simple way to estimate the C/O ratio of the atmosphere is observing the transmission spectra of H$_{2}$O and CH$_{4}$ in the near- to mid-infrared. Both of the systems have zero eccentricities, this means the planet remains at a constant distance from its host star, which avoids orbital-induced thermal response of the planetary atmospheric temperature and hence chemistry \citep{RN539}. Clouds and haze are common in the planetary atmospheres and strongly impact on observations of transmission spectra \citep{RN354,RN713}. Uniform global clouds can obscure the absorption features of prominent chemical species, patchy clouds can mimic high mean molecular weight atmospheres \citep{RN712,RN711} and haze can produce a significant slope at optical passband in transmission spectra \citep{RN429}. According to the theory of \citet{RN429}, assuming a power law for the cross section with wavelength in the form of $\sigma=\sigma_{0}(\lambda/\lambda_{0})^{\alpha}$, the slope induced by Rayleigh scattering is given by 
\begin{displaymath}
		\frac{\mu g_b}{k_B}\frac{d R_p}{d \ln \lambda}=\alpha T_{eq}.
\end{displaymath}
For the cloud-free atmosphere, the Rayleigh scattering is caused by H$_2$ and He, the expected value of $\alpha$ is -4. For the cloud atmosphere, the observable atmosphere would be smaller than pure gaseous atmosphere, so that the amplitude of absorption features would be small or the transmission spectra would be flatten and featureless.

\section{Conclusions}\label{sec:con}

Based on the new photometric data observed by TESS, the YO-1m and the published photometric and RV data from the CDS database and ETD website, we have carried out a re-analysis for the transiting exoplanetary systems WASP-35 and HAT-P-30 by using MCMC technique.  The system parameters and ephemerides of WASP-35 and HAT-P-30 systems have been refined. The refined system parameters are consistent with the previous results with higher precisions. For each system, the uncertainty of the orbital period we obtained is one order of magnitude less than the previous value. Moreover, we find that HAT-P-30b's transits show significant timing variations which cannot be explained with a decaying orbit due to tidal dissipation and the R\o mer effect, while both apsidal precession and an additional perturbing body could reproduce the signal. Based on the refined system parameters, both WASP-35b and HAT-P-30b are suitable for studying the properties of the planetary atmospheres by using transmission spectra.

\begin{acknowledgments}

We appreciate the referee for his/her valuable suggestions and comments, which lead to a obvious improvement to our manuscript. We acknowledge the supports from 1m telescope of Yunnan Observatories. This work is supported by National Natural Science Foundation of China through grants No. U1531121, No. 10873031, No. 11473066 and No. 12003063. We also acknowledge the science research grant from the China Manned Space Project with NO. CMS-CSST-2021-B09. This paper includes data collected by the TESS mission, which is funded by the NASA Explorer Program.

\end{acknowledgments}

%

\vspace{5mm}
\facilities{TESS, YO-1m}


\software{Juliet \citep{RN527}, batman \citep{RN528}, celerite \citep{RN530}, PyMultiNest \citep{RN532,RN531}, emcee \citep{RN263}, Lightkurve \citep{RN521}, TTVFast \citep{RN663}, REBOUND \citep{RN668}}




\bibliography{sample631}{}
\bibliographystyle{aasjournal}



\end{document}